\documentclass[format=acmsmall, review=false, screen=true, authorversion=true]{acmart}
\usepackage{booktabs} 
\usepackage{enumitem}

\usepackage{subcaption}
\usepackage[ruled]{algorithm2e} 
\usepackage{multirow}

\setcopyright{acmlicensed}
\acmJournal{TIIS}
\acmYear{2020} 
\acmVolume{(tbd)} 
\acmNumber{(preprint)}
\acmArticle{1} 
\acmMonth{9} 
\acmDOI{10.1145/3375790}

\begin{document}
\title[CMC for Adaptive Coaching Interactions]{Exploring the Role of Common Model of Cognition in Designing Adaptive Coaching Interactions for Health Behavior Change}

\author{Shiwali Mohan}
\orcid{1234-5678-9012-3456}
\affiliation{%
  \institution{Palo Alto Research Center}
  \streetaddress{3333 Coyote Hill Road}
  \city{Palo Alto}
  \state{CA}
  \postcode{94306}
  \country{USA}}
\email{shiwali.mohan@parc.com}

\renewcommand\shortauthors{Mohan, S.}

\begin{abstract}
Our research aims to develop intelligent collaborative agents that are \emph{human-aware} - they can model, learn, and reason about their human partner's physiological, cognitive, and affective states. In this paper, we study how adaptive coaching interactions can be designed to help people develop sustainable healthy behaviors. We leverage the common model of cognition - CMC \cite{laird2017standard} - as a framework for unifying several behavior change theories that are known to be useful in human-human coaching. We motivate a set of interactive system desiderata based on the CMC-based view of behavior change. Then, we propose PARCoach - an interactive system that addresses the desiderata. PARCoach helps a trainee pick a relevant health goal, set an implementation intention, and track their behavior. During this process, the trainee identifies a specific goal-directed behavior as well as the situational context in which they will perform it. PARCcoach uses this information to send notifications to the trainee, reminding them of their chosen behavior and the context. We report the results from a $4$-week deployment with $60$ participants. Our results support the CMC-based view of behavior change and demonstrate that the desiderata for proposed interactive system design is useful in producing behavior change.

\end{abstract}

%
%


%
%


\maketitle
\section{Introduction}
Mobile health (mHealth) technology, particularly that targeted at obesity and weight-related illnesses, has come a long way since the advent of wearable activity trackers such as \emph{FitBit}. The technology market has been inundated with advanced bio-metric sensors accompanied with data analytics that can monitor and categorize activities performed during the day and innovative interfaces that can help a person gain insights into their health. The stage is set to build mHealth systems that facilitate long-lasting health behavior change. Ongoing research is looking at imparting mHealth systems with human-level coaching capabilities that cancmc help people learn healthy behaviors (such as walking everyday, eating mindfully) and maintain them for a long time. Prior work on mHealth systems has demonstrated how various behavior change theories, generally applied in human-human settings, can be adapted for delivery through technology. This work has explored goal setting theory \cite{consolvo2009goal, konrad2015finding}, rewards and gamification \cite{paredes2013design, helf2015mhealth}, affective forecasting \cite{hollis2015change} etc. Along these lines, our long-term research goal is to develop comprehensive mHealth coaching systems that can continually adapt human-agent interactions to ensure the trainee is making progress towards her health goals.

Our research on AI heath coaching systems contributes to a larger research agenda on design and analysis of \emph{human-aware} AI systems. Usually, when interactive AI or ML systems are designed, little attention is devoted to explicit modeling of the human partner's behavior and decision making. Often, modeling is limited to designing the human-agent interaction patterns. Human-aware AI systems \citep{khampapati2018} seek to address this gap by placing modeling of humans at the center of AI system design. Despite having a tremendous potential, current human modeling methods in AI are extremely limited in their scope \citep{albrecht2018autonomous}. Methods for modeling humans in AI research have originated in game domains where humans are assumed to be other collaborative or opposing agents with \emph{fixed} objective functions. Therefore, proposed formulations are insufficient for representing the dynamic behavior and evolving objective function of humans in the real-world - such as building a healthy habit. By situating human modeling research in the concrete context of health-behavior change, we expect to develop computational models that can be applied to real-world human-agent collaborative tasks. This work contributes to the growing body of work on human-aware AI systems research in diverse domains including transportation \citep{mohan2019acceptable, mohan2019influencing} and robotics \citep{chakraborti2018projection,kim2015inferring}.

Our approach to building human-aware, interactive AI systems is to leverage our understanding of human-human coaching interactions. A crucial observation from effective human-human health coaching is that it is not a singular strategy but a collection of several strategies. A human coach will opportunistically employ different strategies with their trainee. They may create a specific weekly plan for their trainee if they determine that the trainee cannot find time for their health goals, they may set up reminders to make sure that their trainee follows through, or they may provide emotional support if the trainee has been trying but failing to achieve their goals. To design an intelligent coach that can have similar diagnostic and adaptive coaching behavior, it is critical to answer two computational questions. First, what is the causal model of human behavior - what set of factors determine why a target behavior such as \emph{take a walk everyday} does and doesn't occur? Second, what is the space of adaptation in coaching interactions - what are the different ways in which trainee-coach interaction can be varied to ensure progress? By answering these questions, we can develop a comprehensive adaptive strategy for sustainable behavior change that can be easily delivered through a technological medium.

This papers takes an important step towards answering these two questions. To guide our research, we study several behavior change strategies that are shown to be useful in human-human coaching, including goal setting \cite{locke2006new}, setting implementation intentions \cite{gollwitzer2006implementation}, periodic reminding \cite{tobias2009changing}, and leveraging judgments \& attitudes towards healthy behaviors \citep{cafazzo2012design}. Our main conjecture is this - behavior change strategies are effective because they manipulate various aspects of the human cognitive system that underlies all behavior. If we can understand how these strategies influence the cognitive system, we can build a causal model of behavior that is useful in adaptive AI coaching. A cognitive explanation of behavior change also brings several behavior change strategies together, paving the way for characterizing a comprehensive space of adaptation in coaching interactions. To develop our main conjecture, we bring to bear the common model of cognition - CMC \cite{laird2017standard} - that originates from the past $30$ years of cognitive science and AI research on cognitive architectures: Soar \cite{laird2012soar}, ACT-R \cite{anderson2014atomic}, and Sigma \cite{rosenbloom2016sigma}. This paper explores if CMC is useful in reasoning about why behavior-change strategies function, how they can be brought together, and how systems can be designed to leverage them for behavior change. This paper makes the following contributions:
\begin{enumerate}
    \item We provide an integrated account of behavior change by instantiating goal setting, implementation intention setting, reminding, and judgments \& attitudes theories in the common model of cognition (CMC).
    \item We demonstrate that the integrated account of behavior change is useful by motivating a set of desiderata that a comprehensive, adaptive behavior change coach must address.
    \item We discuss the design decisions made for developing an implement mHealth system - \textsc{ParcCoach} and study how they address the desiderata identifed.
    \item We use evidence from a $28$-day, longitudinal, deployment study of \textsc{PARCcoach} (previously reported in \cite{pirolli2016computational}) to demonstrate that the identified desiderata are useful in producing behavior change.
\end{enumerate}

\section{Background \& Related Work}
\label{sec:background}
We begin by briefly reviewing various behavior change theories that are relevant to this paper. Additionally, we review various intelligent interactive systems to characterize previous approaches and situate our research. Finally, we review some publications that are closely related to the work presented here to highlight how this paper contributes to our research agenda.

\subsection{Health Behavior Change Theories}
\label{sec:hbc-theories}
Health-behavior change literature identifies several interventions based on constructs such as self-efficacy \cite{Stacey2015}, implementation intentions \cite{belanger2013meta}, self-affirmations \cite{falk2015self}, motivational interviews \cite{miller2012motivational} etc., that produce large effect sizes in positively influencing health behaviors. Previous work \cite{pinder2018digital} presents a detailed survey of various behavior change theories relevant to behavior change systems. Here, we briefly introduce cognitive theories pertinent to the design of \textsc{PARCcoach} and its assessment in the deployed study.

\subsubsection{Goal Setting}
Having a goal is a crucial cognitive determinant of human behavior and performance \cite{locke2002building}. Success or failure to achieve a goal influences appraisals of other similar goals and motivation to pursue them. Not surprisingly, setting behavioral goals such as \emph{eat more fruits and vegetables} is one of the most promising strategies employed in health behavior coaching \cite{Shilts2004}. However, just setting a random goal will rarely result in behavior change. To be maximally effective, the goals need to be \emph{difficult yet attainable}. They must induce effort for a trainee to be motivating but should not be too difficult that the trainee repeatedly fails at achieving them. They must be \emph{specific}, providing a clear, narrow target and must be \emph{proximal}, mobilizing effort in the near future to ensure commitment and action. As having a good goal is central to behavior change it is not surprising that goal setting interventions are prevalent in mHealth system design and analysis \cite{consolvo2009goal, konrad2015finding, mohan2017designing}.

\subsubsection{Implementation Intentions}
Goals above describe the outcome a trainee would like to achieve after investing time and effort and is committed to. Setting an achievable goal by itself may not guarantee that goal striving behavior will occur. Implementation intentions \cite{gollwitzer2006implementation} serve to fill this so called \emph{intention-behavior gap}. They specify the behavior a trainee will perform in the service of goal achievement and the situational context in which one will enact it. Originally, implementation intentions were defined as \emph{if-then} plans aimed at linking a situation with a goal-directed behavioural response. In a health behavior change context, they may be applied by asking the participant to write plans such as \emph{If I am offered an alcoholic beverage, then I will ...} \cite{armitage2009effectiveness}. However, in the health behavior change practice, a more prevalent form of implementation intention manipulation are questionnaires with prompts to  note the \emph{when} and \emph{where} components of the cues to action \cite{hagger2014implementation}. One key advantage of this method over the original formulation is that specifying the behavior cues (such as time, location) ahead of the intervention reduces the method variance participant-identified cues (\emph{when I am offered a drink}) may introduce. Implementation intentions of the latter type have been widely applied in internet-based interventions using implementation intention \cite{craciun2012facilitating, tam2010planning} and shown to be effective at positively influencing compliance with target behaviors. Our proposed approach here leverages the implementation intentions of the latter type.

\subsubsection{Reminding}
Periodic prompts or messages that are repeatedly delivered to trainees can help them sustain behaviors in service of their health goals \cite{Fry2009}. These prompts can be reminders or brief feedback messages that are communicated to the participants multiple times during the intervention. Reminders serve as external memory aids to initiate or maintain healthy behaviors at appropriate time. Penetration of mobile devices in people's lives has made using reminders for behavior change interventions easier. Previous studies have looked at using text messages to improve medication compliance \cite{martin2005non, downer2005use}, smoking cessation \cite{rodgers2005u}, and diabetes self-management \cite{franklin2006randomized}. Recent efforts have focused on tailoring reminder messages to each individual trainee's needs \cite{woolford2011omg} and just in time interventions that use sensors and analytics to deliver messages at an an appropriate time \cite{thomas2015behavioral}. We present an alternative formulation of reminding that leverages properties of the human cognitive architecture instead of reasoning about the right context using sensors.

\subsubsection{Judgments and Attitudes}
Human motivation to pursue goals is modulated by expectations about the outcome - if the goals/behaviors are perceived to be too difficult, they will not be pursued as the estimated likelihood of success is small (\emph{self-efficacy} is low). Prior work \cite{pirolli2016computational} has shown that such judgments are derived from previous experiences a trainee has with similar goals/behaviors. These constructs have been incorporated in mHealth system and have shown promising results \cite{cafazzo2012design}. Other attitudes \cite{rhodes2003investigating} about health goals and behaviors can be also assumed to be products of prior experiences with them. Affective and instrumental attitude are relevant to this paper. Affective attitude toward a behavior can be interpreted as how much a trainee likes or dislikes performing it. Instrumental attitude represents the degree to which a trainee thinks performing a behavior is `worth-it'. Both of these constructs have been explored in the context of behavior change \cite{tobias2009changing}. In this paper, we present an initial analysis of how these judgements are related to experience and if they contribute to behavior change.

While the efficacy of these theories have been established independently, it is non-trivial to
bring them together into a comprehensive integrated account. The Habit Alternation Model (HAM \cite{pinder2018digital}) is an integrated account of several behavior change theories focussed on explaining how habits are built. The CMC-based account proposed in this paper can be considered a step in the same direction - it is a causal, integrative account of behavior change. However, it is significantly different from HAM. The common model of cognition is a general framework to describe human cognition and has been used to explain a variety of cognitive phenomenon \cite{laird2017standard}. In this paper, we study how this general framework can also be used to understand behavior change theories.
\subsection{Interactive Intelligent Systems}
Developing intelligent coaching algorithms for health behavior change is an intelligent systems problem - not only intelligent methods for adaptation must be studied, implementing effective ways of communicating coaching advice to the human trainee is critical as well.

\subsubsection{mHealth Behavior Change Systems}
Use of technology to affect health behavior change \cite{webb2010using} has been gaining popularity as mobile phones and personal computers become more pervasive. A large number of interventions conducted through technological medium are authored by experts and not personalized to a person's individual needs. The role of technology has been limited to delivering the content in a timely and accessible fashion.

Recently researchers have begun to study how adaptive intelligent systems can aid delivery of behavior change interventions. Prior work \citep{schulman2011intelligent} has studied how conversational agents and dialog systems can be used for motivational interviewing (MI) to promote exercise and healthy eating. \citet{schulman2011intelligent} proposed semantics for MI dialog moves and evaluated the resulting conversational interface with $17$ participants in a laboratory experiment. Each participant had $3$ conversations with the agent and was asked to pretend as if a day had passed between conversations. An expert trained in MI counseling rated a subset of conversations by assessing empathy and fidelity to MI. The agent received high ratings suggesting that the proposed semantics were useful for applying MI technique in conversations.

MI has also been explored from the perspective of virtual agent design \cite{lisetti2013can}. This effort focused on designing a human-like virtual agent to help patients with alcohol addiction. The proposed agent is embodied in a virtual persona and the research has explored how the agent can use facial gestures to display affective empathy and can demonstrate verbal reflexive listening by paraphrasing and summarizing the patient's responses. This agent was evaluated online by recruiting $81$ participants and having them randomly interact with the proposed empathic agent, a non-empathic control agent, or a text-based control system. The results showed that the empathic agent was perceived to be more useful and was more enjoyable to interact with (among other metrics) compared with controls.

Others \cite{Tielman2015} have explored using an ontology-based question system for trauma counseling. The focus of their effort was personalizing the content of the questions to each individual user using an ontology of events. In a laboratory study with $24$ participants, they demonstrated that personalized questions resulted in elicitation of more content from participants in comparison to standard questions.

It is worth noting that evaluation in prior work is largely limited to studying various dimensions \emph{interactivity} and \emph{acceptability} of the proposed AI technology. It is largely silent on whether these methods are useful in producing behavior change in human trainees. Evaluating whether an intelligent agent can produce behavior change is challenging and requires long-term deployment, often for several weeks. Prior work in human-computer interaction \cite{konrad2015finding, hollis2015change} demonstrates how long-term deployments in user populations can be used to measure behavior change and evaluate efficacy of interventions. However, the methods studied provided limited insights about how interactive AI systems must be designed to facilitate health behavior change. Recent work \cite{sidner2018creating} has studied the long-term efficacy of AI technology in the form of dialog agents to support isolated older adults. Although the agents described by \cite{sidner2018creating} do not explicitly target health, it is an example of AI agents supporting wellness goals of their human partners.

Recent work \cite{nahum2015building}  proposed an overarching conceptual vision of how technology can be built to support Just In Time Adaptive Interventions (JITAI) for health-related behavior change. Our approach can be considered an exemplar of such a system. The distal outcome of our approach is to transition the trainee from a sedentary lifestyle to the AHA-recommended aerobic exercise volume per week. The proximal outcome is weekly exercise. The decision point is every day and our adaptive algorithms implement several decision rules based on how experts reason about exercise volume prescription.

\subsubsection{Interactive Tutoring Systems}
There is a long, rich history of using interactive, intelligent technology \cite{Koedinger2013NewOptimization} to support human learning in the classrooms. This line of research on interactive tutoring systems (ITS) \cite{Graesser2001IntelligentDialogue} very closely aligns in motivation and approach to ours. ITS research has looked at measuring and representing what is known by a human user (in domains such as algebra, programming) and adapting a lesson to maximize learning. Efficacy of these model-based interactive, intelligent, technology rivals that of human tutors \cite{Vanlehn2011EducationalSystems}. The efficacy and impact of this interactive technology is a great motivator for incorporating understanding of human cognition in intelligent technology.

ITS research focuses on building up cognitive skills and conceptual knowledge. It makes a critical assumption that the  tasks are performed in focused sessions and all cognitive resources are employed exclusively to the task under consideration. Consequently, the research studies how task-related cognitive skills and conceptual knowledge are learned. Learning for Health-related behaviors is significantly different - it must lead to selection of relevant behavior in ecological settings where other behaviors may compete for cognitive resources. Beyond developing skills and knowledge required for health behaviors, they need to be made more accessible, salient, and valued as compared other tasks a person might be involved in.

This challenge requires us to adopt a more holistic view of behavior, going beyond \emph{knowledge-level} models explored in ITS research. A significant majority of research threads in ITS have looked at augmenting academic learning in classroom environments, often focusing on learning coursework algebra, sciences, and programming. Approaches such as ours will greatly enhance the impact of intelligent tutoring technology by bringing it to the context of learning and sustaining health behaviors.

\subsection{Prior Work}
\label{sec:background-prior-work}
Designing an interactive intelligent system for health-behavior change is a tremendous challenge and requires an inter-disciplinary approach. Over the past few years, we have studied various aspects of this problem resulting in a variety of publications appearing at HCI, Medicine, and AI venues which is indicative of the impact of our theory-driven approach. An early work \cite{konrad2015finding} set up the stage for studying the impact of adaptive mobile technology on developing and sustaining healthier habits through repeated HCI. The paper studied healthy behaviors in three different domains - exercise, meditation, and accessibility. The research effort implemented a quasi-adaptive method to change the schedule of recommended behaviors based on a trainee's reports. The results indicated that adaptation may be useful in sustaining healthy behaviors for a longer time frame.

Following research \cite{du2014efficacy, du2016group} demonstrated that group-based health behavior interventions can be delivered through a mobile medium. While group-based interventions are not directly relevant to the results reported in this paper, this research provides evidence to the hypothesis that a mobile medium can be effectively used to deliver health interventions.

For individually adaptive health interventions, we are developing a \emph{model-based} approach to delivering health interventions through a technology medium. This approach has three prongs: first: developing personalized models of human behavior change; second, adapting interventions, usually administered through papers and diaries in human-human settings, to a mobile medium, and third: developing interactive AI systems that learn and reason with the computational models and deliver individualized interventions through the mobile medium. The central hypothesis to our approach is this - if we can discover precise computational models of how human behavior evolves, we can develop AI methods that exploit these models to guide the human towards better health behaviors.

Along the first prong, \citet{pirolli2016computational} builds a cognitive computational model of self-efficacy. Self efficacy is an individual's confidence that they can perform a behavior successfully. The paper demonstrates that changes in self efficacy can be explained by an individuals past experience with a behavior. \citet{pirolli2017} extend this line of work to model how implementation intentions change behavior and what impact a reminding schedule may have on behavior. Both of these papers use ACT-R, a cognitively plausible computational framework for modeling and aim to model the impact of various interventions on behavior, precisely.

The original paper \cite{konrad2015finding} exemplifies the second prong of our approach. It studies what are the effective ways of delivering adaptive coaching through mobile phone. \citet{Hartzler2016} explore if human participants can interact with the mobile phone app to provide information that an AI algorithm crucially needs to produce reasonable adaptations. Later \citet{springer2018leveraging} study how self-affirmations can be delivered through the mobile phone.

The third prong of our approach focuses on AI system building, leveraging our understanding of the previous two prongs. Prior work \cite{mohan2017designing,mohan2019designing} on the \textsc{NutriWalking} application studies how models of changes in aerobic capability due to exercise can be integrated in an AI adaptive scheduler to achieve behavior change in a relevant population. The model was developed based on how expert physical therapist reason about adapting recommended exercise for a trainee. This line of work not only studies the computational principles that underlie individual adaptation in coaching but also proposes what the trainee-coach interaction should be to support automatic adaptation. While the results from this line of work are encouraging, the work itself is limited in its assumption that behavior change only has a goal setting or skill learning component.

Along the third prong, in this paper, we take a more comprehensive look at interactive AI systems to support health behavior change. We use the common-model of cognition as a guiding framework to bring together several behavior change theories and propose a desiderata for an interactive, intelligent system. We apply the desiderata to design a behavior-change system - \textsc{ParcCoach} and demonstrate its utility in affecting behavior change. The data presented in this paper has been previously reported \cite{pirolli2017} and was used to develop a cognitive model using the ACT-R architecture (prong one in our approach). Along the third prong of intelligent system design, this paper goes into a deeper discussion why certain design choices were made while developing \textsc{ParcCoach} and evaluates the impact of those design choices. Needless to say, both papers share similarities in reporting experiment design and summary statistics. Extending the previous work, this paper views the dataset with a new lens of AI system design and evaluation, advancing our approach to individually adaptive health interventions.

\section{Common Model of Cognition: A Framework for Explaining the Impact of Behavior-Change Theories}
\label{sec:cognitive-architecture-framework}
Using interactive behavior-change strategies in Section \ref{sec:hbc-theories}, an effective coach can influence a trainee to gradually adopt behaviors that align with their goals. One way to view this coaching process is to view coaching interactions as a way to change the cognitive system that underlies a trainee's behavior. If we can explain how various coaching strategies change the cognitive system, we can begin to understand how different strategies can be integrated and how interactive systems can be designed that can bring about a similar change. To explore this further, we bring to bear the common model of cognition \cite{laird2017standard} - CMC and its computational realization in a \emph{cognitive architecture} \cite{langley2009cognitive}. A cognitive architecture's structure defines how computational processes are organized into components and how information flows between components. A central claim of cognitive architecture theory is that the human cognitive system is not an undifferentiated pool of information and processing, but is composed of distinct modules that have specific functionality.

\subsection{The Common Model of Cognition (CMC)}
The CMC is shown in Figure \ref{fig:theory}. It consists of perception and motor control, short-term working memory, long-term declarative and semantic memories, and long-term procedural memory. Working memory is a global transient space within which information is dynamically composed from current perceptions and motor actions. Declarative memory can be considered a composition of two functionalities: semantic memory, a long-term store of facts, concepts, and goals and episodic memory, a long-term store of experiences. Procedural memory contains knowledge about internal and external actions and operates over the contents of working memory. This knowledge includes: which actions are relevant for the current situation, how to select amongst them, and how to execute them. It can be described as pattern-directed invocation of actions and is typically represented as rules with conditions and actions. The conditions specify the information pattern over working memory contents and rule actions modify working memory.

\subsection{Behavior in CMC}
\begin{figure}[t]
    \centering
    \includegraphics[width=0.8\textwidth]{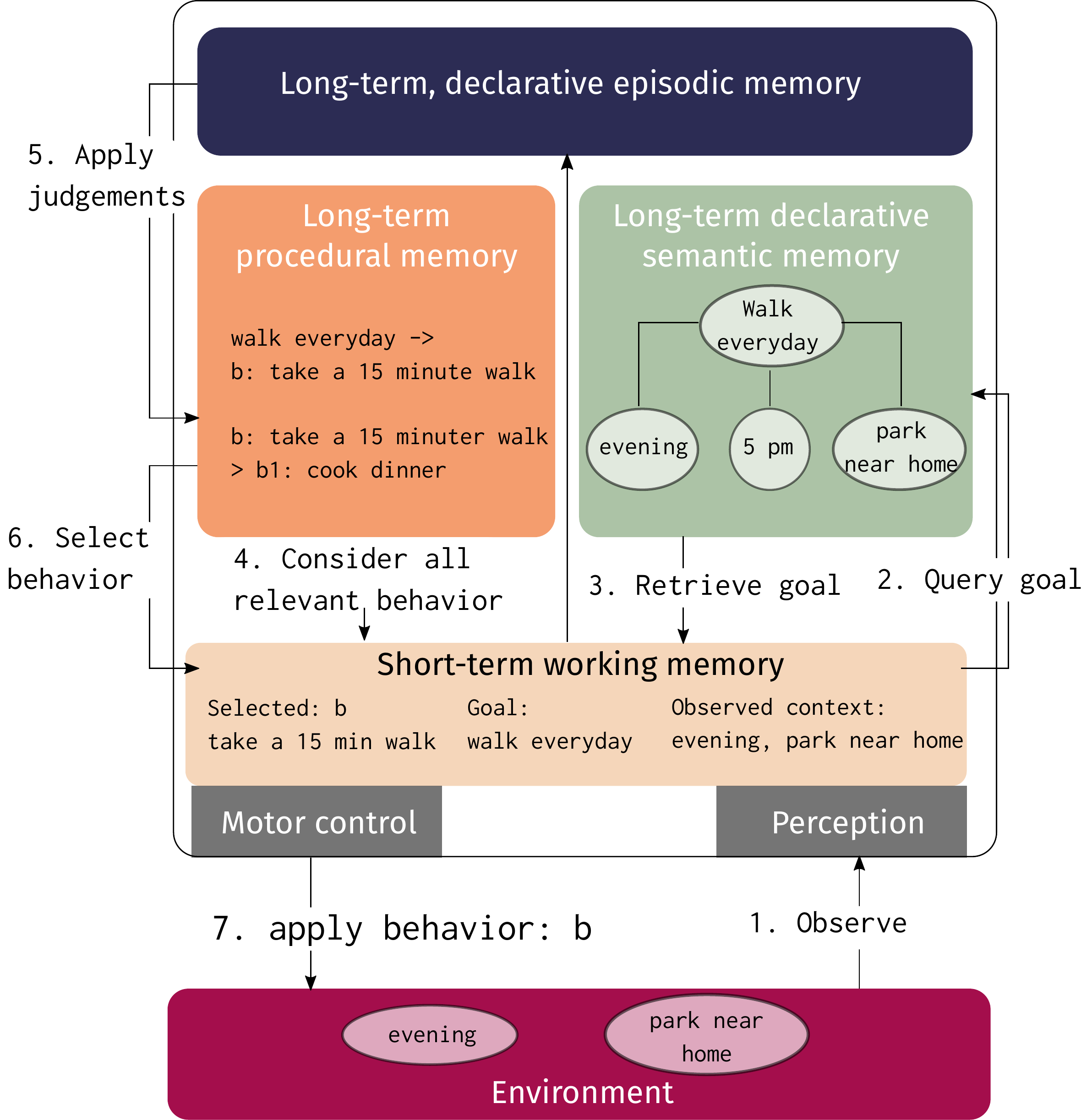}
    \caption{An archetypal cognitive architecture and a model for behavior.}
    \label{fig:theory}
\end{figure}

CMC posits that behavior in the environment arises from a complex interplay of computations in its components and flow of information between them. Consider Figure \ref{fig:theory} again. A target behavior $b$ is performed occurs in the environment as follows (the item numbers correspond to numbered steps in the figure):
\begin{enumerate}
    \item Some critical elements are observed in the environmental context and brought into working memory via a perceptual buffer.
    \item A query to the long-term memory is initiated to see if there are any long-term goals associated with the critical context elements observed.
    \item Long-term semantic memory retrieves a related goal into the working memory.
    \item Various rules in procedural memory match against the contents of working memory to generate a set of all behavior relevant to informational contents of working memory.
    \item Previous experiences are brought to bear for evaluating the utility of various behaviors being considered in the working memory.
    \item A behavior is selected and computational resources are devoted in service of that behavior.
\end{enumerate}

We can map various health behavior-change theories \ref{sec:hbc-theories} on this standard model of the mind. By influencing information processing steps in the CA, health behavior-change theories can cause changes in the behavior of a human cognitive system. For example,
\begin{enumerate}
    \item Goal setting strategy creates a goal (such as \emph{walk everyday}) in long-term semantic memory. Until the goal is achieved, the CMC will occasionally bring this goal into short-term memory so that behaviors in service of this goal can be initiated. Without having an explicit goal, no relevant behavior will be executed and therefore, goal setting crucially impacts behavior change. This impacts step $1$.

    \item Setting an implementation intention creates explicit links between the goal and situational context elements in the environment. Having these explicit links makes the goal accessible when those particular elements are observed in the environment. For example, if \emph{walk everyday} goal is explicitly associated with a specific location \emph{park near my home}, this goal will become accessible whenever a person is at the specific park. On setting an implementation intention, the CMC creates explicit links between their situational context and their goal making it accessible whenever the situational context arises (impacts step $3$).

    \item Frequent reminders incrementally strengthen the links between the context elements and the goal such that the goal becomes more accessible and very easily retrieved whenever the elements are observed in the environment. Reminders that are targeted at strengthening the context-goal links, therefore, can have a powerful impact on influencing behavior (impacts step $3$).

    Previous work \cite{thomas2015behavioral} studies when is the right time to send a message to ensure behavior is executed. CMC posits a formulation of reminder-based interventions that differ from typical formulation that have been previously studied. It suggests that if association between the situational context and behavior is strengthened, the likelihood of behavior increases. Arguably, if the healthy behavior is pegged to the elements of the environmental context instead of a reminding technology, it is much more sustainable and will occur even in the absence of a reminder.

    \item Selecting a specific \emph{how} during implementation intention setting creates an association between the goal and a target behavior in the CA (impacts step $4$). Once the goal become accessible, the CMC begins considering applying $b1$.

    \item However, at any given time a CMC considers multiple goals (such as \emph{cook dinner} along with \emph{walk everyday}) and behaviors relevant to them. If multiple behaviors are under consideration, a CA will evaluate each behavior using various judgments such as self-efficacy, affective attitude, etc. The winning behavior is finally executed. Consequently, A CMC will select the behavior which has been most successful, worth the effort, etc in the past for execution. If an intervention positively influences judgments and attitudes, it increases the likelihood of behavioral performance (impacts step $5 \& 6$).
\end{enumerate}

This CMC-based explanation of behavior change suggests that behavior change theories of goal setting, implementation intention setting, reminding, and judgments \& attitudes are inextricably linked, and therefore, can be brought together. Based on this explanation, we identify the following desiderata for an interactive mHealth coaching system targeted to influencing behavior change. It must initiate interactions to help its trainees:
\begin{enumerate}[label=D\arabic*]
    \item \label{desiderata:1} Create a high-level health goal.
    \item \label{desiderata:2} Associate with the goal a behavior that is of appropriate difficulty and can be performed by the trainee.
    \item \label{desiderata:3} Associate contextual cues from the trainee's environment to the health goal.
    \item \label{desiderata:4} Strengthen the association between context cues and the environment by periodically reminding them of the association.
    \item \label{desiderata:5} Create experiences that positively impact judgments of relevant behaviors to ensure repeated performance.
\end{enumerate}
In the following sections, we describe an implemented system - \textsc{PARCcoach} that is designed to address some of these desiderata. We demonstrate how these ideas can be applied in practice to design an interactive mHealth system for learning new behaviors to combat obesity and weight-related illnesses.

\section{\textsc{PARCcoach}}
\label{sec:parccoach}
\begin{figure*}[t!]
    \centering
    \includegraphics[width=1\textwidth]{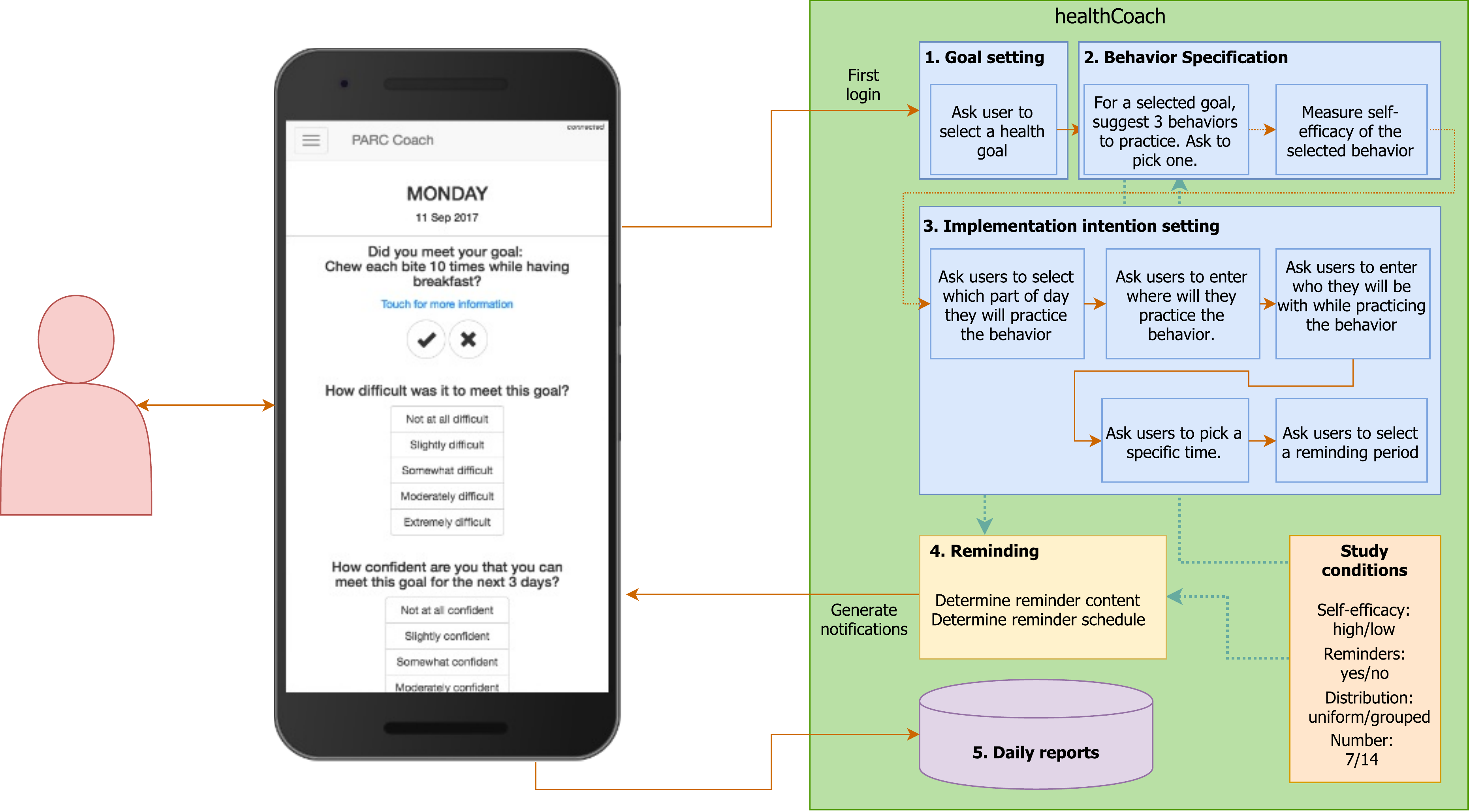}
    \caption{PARCoach helps trainees select a relevant goal, set an implementation intention to perform a behavior, and periodically reminds them of the behavior as well as the situation context in which the behavior must be performed. It additionally tracks behavior compliance through self-reports.}
    \label{fig:coach}
\end{figure*}
PARCoach is a health web-application developed using Meteor \footnote{https://www.meteor.com/} that can be deployed to smartphones and tablets. It lets people track their health behaviors by reporting their success or failure at performing a behavior daily as shown in Figure \ref{fig:coach}. It is designed to be interactive; it can pose questions to the trainee through a pop-up dialog box as well as send notifications to their devices prompting them to return to the app. In the following sections, we describe how we use these interactions to implement theory-guided coaching strategies. PARCoach leads the trainee through the following phases (the numbers here map to those in Figure \ref{fig:coach}):

\subsection{Goal Setting}
\label{sec:goal-setting}
As we discussed earlier, selecting a relevant goal is crucial to any behavior change (desideratum \ref{desiderata:1}). The first time a trainee registers for \textsc{PARCcoach}, it lets the trainee pick a goal for making their lifestyle healthier. We worked with an expert human coach to identify a set of goals that are conducive to be used in a setting such as ours. The criteria was that these goals should not require the trainee to learn a new, complex skill (such as new strength exercises) and that a trainee should be able to monitor their behavior and report success or failure with relative ease. This was done to minimize the need for other coaching strategies not being studied such as skill training, diagnosis and problem solving, etc. that are required for learning a new behavior. For this study, \textsc{PARCcoach} let trainees select from $3$ goals: \emph{eating slowly and mindfully}, \emph{walking everyday}, and \emph{eating more fruits and vegetables}. These health goals are very relevant to modern lifestyle where a significant population is engaged in sedentary office work and does not have enough time to plan and focus on nutrition. We chose to let trainees make selection instead of assigning them a goal. This ensures that the trainees are highly motivated to use \textsc{PARCcoach} daily because they are pursuing a goal that they think will be valuable for them.

\subsection{Behavior Specification}
\label{sec:behavior-specification}
Once the trainee has selected a goal, \textsc{PARCcoach} moves to the next phase of coaching which involves picking a specific \emph{target} behavior. As described in the previous section, selecting a target behavior creates explicit association with the health goal. This step is an important component of both the goal-setting theory as well as the implementation intention setting theory. Additionally, having a target behavior to perform everyday makes it easy to monitor and report success or failure. With discussions with an expert human coach, we developed a list of behaviors that can achieve each high-level goal goal. Examples of selected target behaviors is shown in Table \ref{tab:goals_behaviors}.

\begin{table}[]
\centering
\begin{tabular}{@{}ll@{}}
\toprule
Goals & Behaviors \\ \midrule
\emph{Eat slowly and mindfully} & Chew each bite $10$ times \\
 & Take $30$ minutes to consume a meal \\
 & Chew each bite until it is liquified \\
\emph{Walk everyday} & Stretch for 5 minutes and walk for 15 minutes \\
 & Walk for 10 minutes \\
 & Stretch for 10 minutes and walk for 45 minutes \\
\emph{Eat fruits and vegetables} & Double the serving of your favorite vegetable \\
 & Have a salad. \\
 & Try a new vegetable \\ \bottomrule
\end{tabular}
\caption{Examples of target behaviors for $3$ goals}
\label{tab:goals_behaviors}
\end{table}

Goal setting theory suggests that an ideal coach should help a trainee select a target behavior that is \emph{difficult} yet \emph{attainable} to practice (desideratum \ref{desiderata:2}). To do so, a coach must a) know how difficult these behaviors are relative to each other as well as b) what difficulty level is appropriate (\emph{attainable}) for the trainee. Our approach to answering these questions is below.
\begin{figure}
    \begin{subfigure}{0.32\linewidth}
        \centering
        \includegraphics[width=1\textwidth]{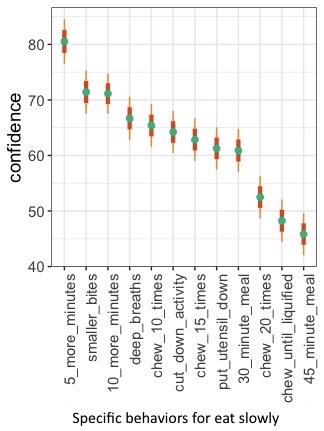}
    \end{subfigure}
     \begin{subfigure}{0.32\linewidth}
        \centering
        \includegraphics[width=1\textwidth]{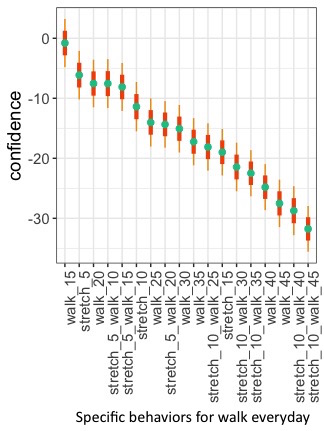}
    \end{subfigure}
    \begin{subfigure}{0.32\linewidth}
        \centering
        \includegraphics[width=1\textwidth]{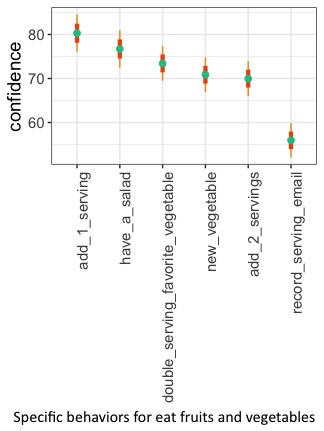}
    \end{subfigure}
    \caption{Coefficient plots representing self-efficacy for target behaviors for \emph{eat slowly and mindfully}, \emph{walk everyday}, and \emph{eat fruits and vegetables} health goals. The green points represent the linear regression coefficients, red thick bars $50\%$ confidence interval, orange thin bars $95\%$ confidence interval.}
    \label{fig:self-efficacy-measurement}
\end{figure}

\subsubsection{a) Measuring Target Behavior Difficulty}
\label{sec:behavior-difficulty}
It is easy to quantify difficulty of some behaviors such as \emph{walking n minutes}, where difficulty is directly proportional to the duration of the walk. Consequently, a $15$ minute walk is easier than a $45$ minute walk. However, quantifying relative difficulty for other behaviors such as \emph{chew each bite 10 times} and \emph{chew each bite until liquefied} is not trivial. We estimated the difficulty of target behaviors through a survey study on MTurk.

\paragraph{Materials}:
In consultation with a human health and lifestyle coach, we generated a collection of target behaviors for each goal: $12$ for \emph{eat slowly and mindfully}, $18$ for \emph{walk everyday}, and $6$ for \emph{eat fruits and vegetables}. We created a survey questionnaire with two measurement blocks: \emph{self-efficacy} and \emph{perceived difficulty} along with questions for a participants demographic information. Each block had $12$ target behaviors randomly selected for each participant for measurement as well as one validation question that was displayed before and after them. Each target behavior was displayed on a separate page with along with a measurement question based on which block the participant was in:
\begin{itemize}
    \item \textbf{Self-efficacy}: \emph{How confident are you that you can perform this behavior everyday for the next 7 days?} The responses were measured on a numeric scale of range \emph{Not confident at all} ($0\%$) to Very confident ($100\%$).
    \item \textbf{Difficulty}: \emph{How difficulty is it for you to perform this behavior everyday for the next 7 days?} The responses were measure on a numeric scale of range \emph{Very easy} ($1$) to \emph{Very difficult}  ($10$).
\end{itemize}

\paragraph{Participants}
The survey had $1523$ adult participants with $1393$ who completed the survey. Their age ranged $18$ - $76$ (mean = $34.59$, sd = $11.41$) with $890$ of them identifying as female, $497$ as male, and $6$ as non-binary. $70.6\%$ of the participants were $21$-$40$ years old, $77.8\%$ had a sedentary lifestyle, and most were actively involved in or thinking about healthier lifestyles. Each participant was paid x for taking the survey.

\paragraph{Data Analysis}
After removing noise from careless answering using the validation questions, each target behavior received $175$ - $200$ measurements for each question above. We fit a mixed-effects linear model to this data of the form $y = \beta x + \gamma z + \epsilon$, where $y$ is the self-efficacy/difficulty ratings vector, $x$ is a fixed effect vector corresponding to target behaviors, and $z$ is a random effect vector corresponding to a participant. $\beta$ and $\gamma$ are coefficients and $\epsilon$ is the error term. Participants were included as random effects in this model to account for individual differences in interpretation of the scale.

\paragraph{Results}
The results from this model are in Figure \ref{fig:self-efficacy-measurement}. The green points represent how participants measured their confidence in being able to regularly perform behaviors. The red and orange bars represent $50\%$ and $95\%$ confidence intervals. The behaviors in each group are ordered in decreasing confidence (or increasing difficulty). The mean self-efficacy and mean difficulty measurements for target behaviors were strongly correlated (Pearson's correlation coefficient = $-0.9634162$).

This measurement model recovers the correct ordering of behaviors that have quantifiable difficulty as evidenced by the ordering of \emph{walking} behaviors in Figure \ref{fig:self-efficacy-measurement}, center. This suggests that people's judgment of self-efficacy and difficulty can be used to obtain a difficulty ordering of behaviors. Further, as self-efficacy and mean difficulty measurements are correlated, they can be used interchangeably.

\subsubsection{b) Determining Appropriate Target Behavior}
\label{sec:appropriate-specific-behavior}
To answer question b) i.e., what difficulty level is appropriate for behavior change coaching, we created an experiment condition. For each goal, we created two sets of $3$ target behaviors: \emph{easy} and \emph{hard}. These sets were created using the analysis above selecting the three least and three most difficult behaviors in the ordering. Each participant was randomly selected into these groups. Our findings from this randomization is discussed later in Section \ref{sec:evaluation}.

Based on which goal a trainee selected and which experimental group they were assigned to, \textsc{PARCcoach} presented them with $3$ target behaviors and prompted them to pick one for practicing for next 4-weeks. Once the trainee makes a selection, we also measured their self-efficacy on the selected target behavior. We adapted the numeric self-efficacy scale to an ordinal scale with levels: \emph{not at all confident}, \emph{slightly confident}, \emph{somewhat confident}, \emph{moderately confident}, and \emph{very confident}. The scale was adapted was done to make answering this question on a smartphone screen convenient.

\subsection{Implementation Intention Setting}
\label{sec:implementation-intention}
We adopted a formulation of implementation intention setting \cite{hagger2014implementation} for \textsc{PARCcoach} design which creates associations between predetermined situation context cues such as time, location etc. with a target behavior. Through a series of prompts, PARCCoach helps people identify cues for the following situational components:
\begin{enumerate}
\item \emph{Time of day/meal}: For the \emph{walking every day} target behaviors, PARCCoach helped the trainees pick a part of day from morning, afternoon, evening, or night when they will practice it. For the \emph{eating slowly and mindfully} and \emph{eat fruits and vegetables}:  PARCCoach helped trainees pick a meal to practice the behavior from breakfast, lunch, and dinner.
\item \emph{Location}: PARCCoach asked trainees to specify where they are most likely to go for a walk or have the meal selected in the previous step using free text. Answers to this included string such \emph{home}, \emph{in park nearby} etc.
\item \emph{Person}: PARCCoach, then, helped trainees specify the social context of the behavior by identifying who they are most likely to practice the behavior with.
\item \emph{Specific time}: Then, the trainees picked a specific time when they are most likely to practice their selected behavior.
\item \emph{Reminder period}: Finally, PARCCoach asked the trainees to specify when they would like to be reminded of the behavior. The trainees were given an option to be reminded $15$, $30$, $45$, or $60$ minutes before the time scheduled for practice. PARCCoach uses this information to generate reminders as described in the next section.
\end{enumerate}
Through these steps, every trainee identified the context in which they will practice their selected behavior. For the purposes of this study, this context was set only once when the trainees first logged into PARCCoach. As the context is defined for one part of day/meal in a day, trainees were expected to monitor and report success and failure only for that one instance every day. Setting an implementation intention in the way proposed here addresses the design desideratum \ref{desiderata:3} and set up the stage for a novel reminding strategy as described below.

\subsection{Reminding}
For addressing desideratum \ref{desiderata:4}, PARCCoach uses the information provided in the previous phase to generate personalized, time-sensitive reminders. Reminders in our system have the following three components:
\begin{enumerate}
    \item \emph{Message}: The goal of reminding is to strengthen the association links between the goal, the target behavior, and the situational context. An example reminder message from PARCCoach is - \emph{"Please remember to chew each bite 10 times while having dinner, at: home, with: with my husband"}. In this example, the trainee had selected to practice the habit of \emph{chewing each bite 10 time} during dinner and had provided \emph{home} and \emph{with my husband} as input for the location and person components during implementation intention setting.
    \item \emph{Active period}: The reminders are sent as a notification to the trainee's mobile device at the time they had indicated they would like to be reminded at. This time was calculated by subtracting the reminder period above from specific time. The notification brings the trainee into the application where a dialog box with the message is displayed. The dialog box remains on the screen until the trainee acknowledges the message by clicking `OK'. The reminder remains active until the specific time. If the trainee acknowledges the reminder after the specific time, they don't see the reminder message.
    \item \emph{Days vector}: This vector represents the days on which a reminder will be scheduled to be delivered to the trainee. With reminders, a concern is that if the reminders are too frequent they become less salient and therefore, may cease to be effective. To study this impact, we randomly controlled the number of reminders a trainee would see. Every trainee received either $7$ or $14$ reminders in the study, with no more than $1$ in a day. The reminders can have varying distribution. \emph{Uniform} distribution uniformly spaces the reminders while \emph{massed} distribution uniformly spaces pairs of reminders as shown in Table \ref{tab:reminders-set}.
\end{enumerate}

\begin{table}[]
\centering
\begin{tabular}{@{}ccc@{}}
\toprule
number & distribution & reminder days \\ \midrule
7 & uniform & {[}4,8,12,16,20,24,28{]} \\
 & massed & {[}3,4,11,12,19,20,27{]} \\
14 & uniform & {[}2,4,6,8,10,12,14,16,18,20,22,24,26,28{]} \\
 & massed & {[}3,4,7,8,11,12,15,16,19,20,23,24,27,28{]} \\ \bottomrule
\end{tabular}
\caption{Reminder schedules for different conditions.}
\label{tab:reminders-set}
\end{table}

It is noteworthy that this formulation of reminding is significantly different from other \emph{just in time} formulations \cite{thomas2015behavioral}. In those, the focus is on using sensors and analytics to determine the correct time to give a reminder. Predicting when is the right time to send a reminder can be computationally challenging. Here, we strengthen the association between context and behavior through reminders. Here we rely on the CMC to retrieve appropriate goals and behaviors when the right context arises in the trainee's environment.

\subsection{Daily reporting}
\label{sec:daily-reporting}
The trainees were expected to login everyday and report success and failure at performing the target behavior in the situational context as set above. An example of the reporting screen is shown in Figure \ref{fig:coach}. The participant can select if they practiced the behavior by clicking on the $\checkmark$ if successful or $\times$ if failed. Through this daily reporting, PARCCoach can observe the experience it creates through its coaching interactions. To study if these experiences have a positive impact on a trainee's evolving judgments and attitudes, PARCCoach also measured the following quantities:
\begin{enumerate}
    \item \emph{Difficulty}: As in the survey described previously, this quantity represents how difficult the trainee perceives the behavior to be. It was measured using the question \emph{How difficult was it?}. The responses were obtained on an ordinal scale ($1$-$5$): \emph{not at all difficult} ($1$), \emph{slightly difficult} ($2$), \emph{somewhat difficult} ($3$), \emph{moderately difficult} ($4$), and \emph{extremely difficult} ($5$).
    \item \emph{Self-efficacy}: As discussed earlier, this quantity measures how likely the trainee expects to be successful at performing a behavior. As in the survey, it was measured through the question \emph{How confident are you that you can meet this goal for the next 3 days?}. The responses were on an ordinal scale ($1$-$5$): \emph{not confident at all} ($1$), \emph{slightly confident} ($2$), \emph{somewhat confident} ($3$), \emph{moderately confident} ($4$), and \emph{extremely confident} ($5$).
    \item \emph{Affective attitude}: This quantity represents the affective connotation of a behavior i.e. how much a person likes or dislikes performing a behavior. It was measured through the question \emph{How inclined are you to do this every day?}. The responses were on an ordinal scale ($1$-$5$): \emph{not at all keen} ($1$), \emph{somewhat keen} ($2$), \emph{moderately keen} ($3$), \emph{quite keen} ($4$), and \emph{very keen} ($5$).
    \item \emph{Instrumental attitude}: This quantity captures the degree to which the trainee feels it was \emph{worth it} to perform a behavior. It was measured using the question \emph{Was doing it worth the effort?}. The responses were on an ordinal scale ($1$-$5$): \emph{much more effort than benefit} ($1$), \emph{some more effort than benefit} ($2$), \emph{almost the same effort as benefit} ($3$), \emph{some more benefit than effort} ($4$) and \emph{much more benefit than effort} ($5$).
\end{enumerate}

Note that trainees cannot go and `back-report' on days past thereby minimizing late and potentially erroneous reporting.

\section{Experimental Evaluation}
\label{sec:evaluation}
Until now, we have proposed an integrative account of behavior change based on the CMC which brings together several behavior change theories in a single framework.  These theories include: goal setting, implementation intention, reminding, and judgment \& attitudes. We used this integrated account to motivate design desiderata for behavior change coaching systems. \textsc{PARCcoach} described in the previous section implements a subset of the desiderata identified. In this section, we evaluate the CMC-based integrated account by studying the empirical evidence obtained through deployment of \textsc{PARCcoach} to $60$ participants and recording their behavior for over $28$ days.

\subsection{Hypotheses and Evaluation Plan}
 To guide our empirical analysis, we first delineate our hypotheses that follow by assuming the CMC-based integrative account of behavior change:
\begin{enumerate}
    \item Setting explicit, specific goals improves behavior compliance.
    \item Operationalizing goals by associating environmental, situational context to a target behavior should improve behavior compliance.
    \item Reminders that strengthen the association between situational, contextual cues should improve behavior compliance.
    \item Individual, experiential difference in evaluating self-performance should influence behavior compliance.
\end{enumerate}
Of these, $1$ and $2$ have been studied extensively in previous goal setting and implementation intention literature. CMC suggests that implementation intentions will increase the likelihood of behavior compliance above what is achieved by goal setting alone. The effect on compliance is additive because these interventions positively affect different parts of the CMC. Goal setting creates a relevant goal structure in long-term memory that is intermittently retrieved in working memory to evaluate if efforts should be invested in its pursuit, implementation intentions associate situational context cues with the goal increasing the likelihood that it will be retrieved. The additive effect of implementation intentions on goal setting has been validated empirically \cite{sheeran1999implementation}. From the perspective of designing mHealth systems, $1$ suggests that desideratum \ref{desiderata:1} is useful and $2$ suggests that \ref{desiderata:3} is useful.

In our evaluations, we incrementally study hypotheses $3$ and $4$ by observing the impact of \textsc{PARCcoach} on behavior change in study participants as enumerated below. In studying these hypotheses, we also evaluate the usefulness of desiderata \ref{desiderata:2}, \ref{desiderata:4}, \& \ref{desiderata:5}.
\begin{enumerate}[label=H\arabic*]
   \item \label{hypothesis:difficulty} Baseline ease of target behaviors positively influences behavior compliance. CMC-based integrated account suggests that people who begin practicing easier behavior are likely to have more successes and eventually, stick with the behavior longer. People who begin practicing hard behaviors may drop of earlier. If this hypothesis is correct, then system desideratum \ref{desiderata:2} is useful in designing effective adaptive mHealth systems.
   \item \label{hypothesis:frequency} Frequency of reminders positively influences behavior compliance. CMC-based account suggests a reminding strategy that is unique in mHealth systems - it relies on the CMC to retrieve right goals at the right time. We expect that this strategy, like other just-in-time reminding strategies, will improve behavior compliance. Additionally, because it works by associating elements of the situational, environmental context with a target behavior, this positive impact will be seen even on days when there are no reminders. If this hypothesis is correct, then desideratum \ref{desiderata:4} is useful.
   \item \label{hypothesis:reminding-additive} Reminding improves behavior compliance above and beyond what is explained by changes in difficulty or estimate ease of a behavior. Because they positively affect different aspects of the CMC, their effects are additive. As with hypothesis \ref{hypothesis:frequency}, this hypothesis also is relevant for evaluating desideratum \ref{desiderata:4}.
   \item \label{hypothesis:value} Value judgments over past experiences influence behavior compliance. CMC asserts that before a behavior is executed, it is evaluated amongst a set of competing behaviors. If the behavior is evaluated highly, it is more likely to be complied to. This hypothesis is relevant to desideratum \ref{desiderata:5}.
\end{enumerate}

\subsection{Method}
We recruited $85$ participants through student email lists of Universities of Michigan, Ann Arbor and California, Santa Cruz as well as an internal list of participants who we had surveyed previously and had expressed an interest in participating in health-related studies. Of these, only $64$ completed the pre-survey, downloaded and signed up on \textsc{PARCcoach} on their smartphone. Compensation was pro-rated and these participants received \$$20$ for finishing these 2 steps. Of these, $60$ participants (mean age $31.59 \pm 9.89$ years; median $30$ years) participated in the $4$-week study and completed the post-survey and received an additional \$$30$ for finishing the study. Four participants were dropped from the study as they did not complete the post-survey and ceased communication with the research team.

The study had a partial-factorial design. The participants took a pre-survey in which they provided us with some demographic information as well as consented to participating in the study. Participants who completed the pre-survey were whitelisted on the \textsc{PARCcoach} server and were asked to signup for it. As participants signed up for PARCoach, they were randomly sorted into different conditions. First, they were asked to choose between the three types of goals: \emph{eating mindfully}, \emph{eating more fruits and vegetables}, and \emph{walking everyday}. Then, they were randomly sorted into target behavior difficulty \emph{high} and \emph{low}. Based on this condition, they were shown a group of relevant target behaviors to choose from. After making a selection of a target behavior to practice, the participants set their implementation intention as in Section \ref{sec:implementation-intention}. Participants were sorted  into reminder \emph{yes} and \emph{no}. Those who were sorted into \emph{yes} were then randomly assigned to \emph{uniform} or \emph{massed} distribution and then to \emph{7} or \emph{14} reminding instances. Participants were blind to which study conditions they were assigned to. The sample size was determined through a power analysis for goodness-of-fit tests with expected effect size $= 0.5$, error probability $= 0.05$, and degree of freedom $= 9$ ($10$ conditions in the experiment) which gave us a sample size of $63$. Distribution of participants across the $10$ study conditions are shown in Table \ref{tab:conditions}.

\begin{table}[h]
\centering
\begin{tabular}{@{}ccccc@{}}
\toprule
\multicolumn{1}{l}{behavior difficulty} & \multicolumn{1}{l}{reminders} &  \multicolumn{1}{l}{reminder distribution} & \multicolumn{1}{l}{reminder count} & \multicolumn{1}{l}{n} \\ \midrule
high & no & none & & 5 \\
high & yes & uniform & 7 & 7 \\
high & yes & uniform & 14 & 6 \\
high & yes & massed & 7 & 5 \\
high & yes & massed & 14 & 4 \\
low & no & none & & 7 \\
low & yes & uniform & 7 & 7 \\
low & yes & uniform & 14 & 6 \\
low & yes & massed & 7 & 7 \\
low & yes & massed & 14 & 6 \\ \bottomrule
\end{tabular}
\caption{Number of participants in each condition}
\label{tab:conditions}
\end{table}

\subsection{Data Analysis}
Through the application, we collected the following daily data about the participants' behavior:
\begin{itemize}
    \item Daily behavior reports: Each day, participants report if they completed the target behavior or not. If they do not make any reports, an absence is recorded. Thus for each of $63$ participants, we have $28$ observations about their behavior. Daily observation is made through a nominal variable with values: \emph{success}, \emph{failure}, or \emph{absent}. From these reports, we extract our main dependent variables for each day:
    \begin{enumerate}
        \item a binary variable \emph{reported} that has the value $1$ if a participant made \emph{success} or \emph{failure} report and a value $0$ if they were absent; and
        \item a binary variable \emph{completed} that has the value $1$ if a participant made a \emph{success} report and $0$ otherwise.
    \end{enumerate}
    \item Judgements: In addition to behavior reports, we measured participants perceptions of the target behavior's difficulty, their \emph{self-efficacy} at performing that behavior daily, and their \emph{affective} and \emph{instrumental} attitudes about performing the behavior as described in Section \ref{sec:daily-reporting}.
\end{itemize}
In addition to these daily reports, we also conducted pre- and post-surveys for collecting participants' weight, weekly physical activity level (duration and frequency \cite{Booth:2003ui}), and eating behaviors in terms of time taken to eat a meal and servings of fruits and vegetables consumed. Additionally  we asked participants to report their readiness to change their exercising and eating behaviors on an ordinal scale mapping the $5$ stages of behavior change according the transtheoretical model \cite{prochaska2013transtheoretical} going from pre-contemplation, contemplation, preparation, action (i.e., behavior adopted for $<6$ months) to finally maintenance (i.e., behavior adopted for $>6$ months). Finally, in our post-survey, we administered the $10$-item System Usability Scale (SUS) \cite{sauro2011practical} to evaluate the usability of PARCcoach.

Using these data, we conducted the following analyses to evaluate the impact of various experimental and incidental conditions.
\begin{itemize}
    \item In Section \ref{sec:overview}, we summarize choices made by participants in the study and their overall reporting behavior.
    \item In Section \ref{sec:baseline_behavior}, we characterize baseline reporting and successful completion of target behaviors. As we have multiple observations from a single participant, we employ mixed-effects logistic regression models. The model is of the form $Pr(y) = 1 / ( 1 + e^{-(\alpha + \beta x + \gamma z + \epsilon)})$, where $Pr(y)$ is the probability of dependent variable (\emph{reported} and \emph{completed} in our case), $x$ is a fixed-effect vector corresponding to the number of days the participant as been in the PARCcoach system, and $z$ is a random effect vector corresponding to a participant. $\alpha$ is the intercept and $\beta$ and $\gamma$ are coefficients and $\epsilon$ is the error term. Participants are modeled as random effects in these models under the \emph{random intercept} assumption which encodes that some variance in responses is due to each participant being different. We report $R^2m$, the proportion of variance in $y$ explained by measured, independent variables as well as $R^2c$, the proportion of the variance that is explained by the complete model including the random individual-level effects.
    \item In Section \ref{sec:analysis_difficulty}, we evaluate hypothesis \ref{hypothesis:difficulty} - how does task difficulty impact reporting and completion of target behaviors. For this analysis, we extend the mixed-effects models above to include two different measures of difficulty - population-level difficulty measurement as estimated from the MTurk study and self-evaluation of difficulty as measured through the self-efficacy scale.
    \item In Section \ref{sec:analysis_reminders} we evaluate hypothesis \ref{hypothesis:frequency}. Here, we analyze the impact of reminding frequency and distribution on reporting and completion by extending the mixed-effects model to include these independent variables.
    \item In Section \ref{sec:analysis_difficulty_reminders}, we evaluate \ref{hypothesis:reminding-additive}. We extend the mixed-effects logistic regression to study the joint impact of task difficulty and reminding on reporting and compliance.
    \item In Section \ref{sec:analysis_judgments}, we consider \ref{hypothesis:value}. First, we study how various judgement measurements - perceived difficulty, self-efficacy, instrumental attitude, and affective attitude - are influenced by a trainee's experience as measured through reports - success, failure, or absent. For this analysis, we developed mixed-effects linear regression models of the form $y = \alpha s + \beta f + \gamma a + \eta z + \epsilon $. $y$ is the value judgment, $s$ is the number of successes experienced until the judgment measurement was made, $f$ is the number of failures, $a$ is the number of absent reports, and $z$ is a random effect vector corresponding to a participant. $\alpha$, $\beta$, $\gamma$ are coefficients to be estimated and $\epsilon$ is the noise parameter. As previously, participants are modeled as random effects to encode variance due to individual differences. Next, we study the influence of measured value judgement on probability of successfully completing a target behavior. For this analysis, we rely on mixed-effects logistic models with success, failure, and absent modeled as dependent variables.
\end{itemize}
As noted in \ref{sec:background-prior-work}, in a previous work \cite{pirolli2017}, we analyzed a part the dataset from this study to develop a cognitive model of behavior change. This paper focuses on the design of mHealth systems and uses the dataset to justify the desiderata identified in Section \ref{sec:cognitive-architecture-framework}.

\subsection{Results}

\subsubsection{Overview}
\label{sec:overview}
\begin{figure}[h!]
    \centering
    \begin{subfigure}{0.4\linewidth}
        \includegraphics[width=1\textwidth]{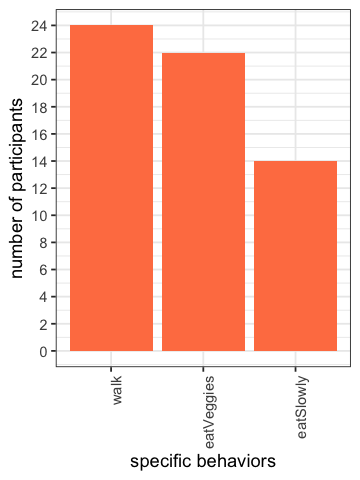}
        \caption{Distribution of participants across various goals in PARCoach}
        \label{fig:summary_goals}
    \end{subfigure}
    \hfill
    \begin{subfigure}{0.4\linewidth}
        \includegraphics[width=1\textwidth]{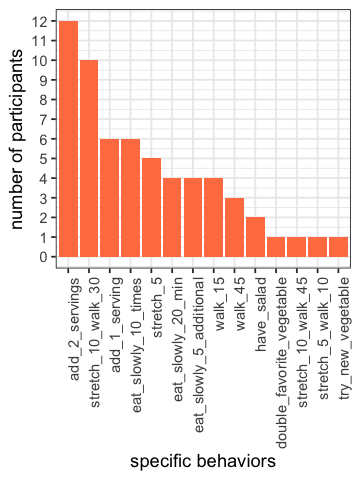}
        \caption{Distribution of participants across various target behaviors.}
        \label{fig:summary_behaviors}
    \end{subfigure} \\
    \begin{subfigure}{0.4\linewidth}
        \includegraphics[width=1\textwidth]{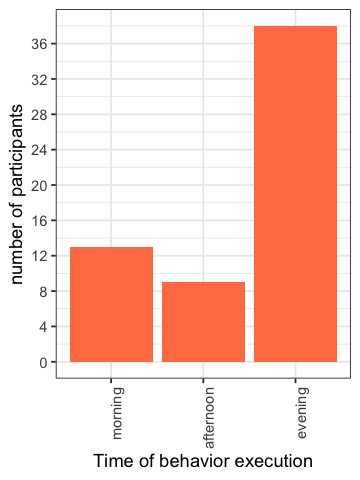}
        \caption{Distribution of participants across time of behavior execution}
        \label{fig:time_behaviors}
    \end{subfigure}
    \hfill
    \begin{subfigure}{0.4\linewidth}
        \includegraphics[width=1\textwidth]{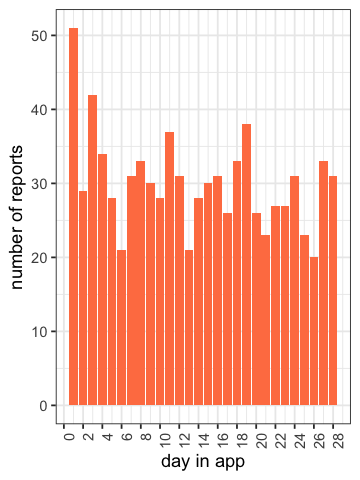}
        \caption{Distribution of participants across time of behavior execution}
        \label{fig:reports}
    \end{subfigure}
    \caption{Summary of participant choices of goals, target behaviors, and time of behavior execution}
\end{figure}

We begin by briefly summarizing choices made by various participants and their overall reporting behavior in the study. Participants self-selected into the three health goal groups with $24$ participants(in Figure \ref{fig:summary_goals} choosing to practice \emph{walk}, $22$ choosing to practice \emph{eating more vegetables}, and $12$ choosing to practice \emph{eating more mindfully}. If we look at choices made for target behaviors in Figure \ref{fig:summary_behaviors}, we see that most people selected \emph{adding 2 servings of vegetables} and \emph{stretching 10 minutes and walking 30 minutes}. Together these histograms show that people selected a variety of healthy behaviors to practice for the length of the study. In Figure \ref{fig:time_behaviors}, we see that most people preferred to practice behaviors in the evening. Figure \ref{fig:reports} tabulates the number of reports made every day in the application. We see that $20 - 30$ participants report every day even towards the end of the study.

\newpage

\subsubsection{Baseline Behavior}
\label{sec:baseline_behavior}
To understand the baseline reporting behavior in our application, we fit a mixed-effects logistic regression models with \emph{day} in the app as the independent variable and dependent variables corresponding to if any reports were made (\emph{reported}) and if successful reports were made (\emph{completed}). Table \ref{tab:baseline} summarizes both these models. We see that the number of days in the app has a small but significant, inverse relationship with making reports. Overall, participants tended to \emph{drop-out} of the application as time progressed. The negative correlation of day in app with successful completion is not as significant.

\begin{table}[h]
\centering
\def\arraystretch{1.2}
\begin{tabular}{lcc}
\toprule
\multicolumn{3}{c}{MODEL I}\\
\midrule
dependent $\rightarrow$ & \multirow{2}{*}{\emph{reported}} & \multirow{2}{*}{\emph{completed}} \\
independent $\downarrow$ &  &  \\ \midrule
\emph{day} &  $-0.032^{***}$  &  $-0.013.$\\ \midrule
$R^2_m$ & $0.011$  & $0.002$ \\
$R^2_c$ & $0.468$ & $0.478$ \\\bottomrule
\end{tabular}
\caption{Mixed-effects regression models for reports made and behavior successfully completed as a function of day in the app. p-value significance codes:  $^{***} 0.001, ^{**} 0.01, ^{*} 0.05, . 0.1$ }
\label{tab:baseline}
\end{table}

\subsubsection{Difficulty of Target Behaviors}
\label{sec:analysis_difficulty}
\begin{figure*}[h]
    \centering
    \begin{subfigure}[t]{0.45\linewidth}
        \centering
        \vskip 0pt
        \includegraphics[width=0.7\textwidth]{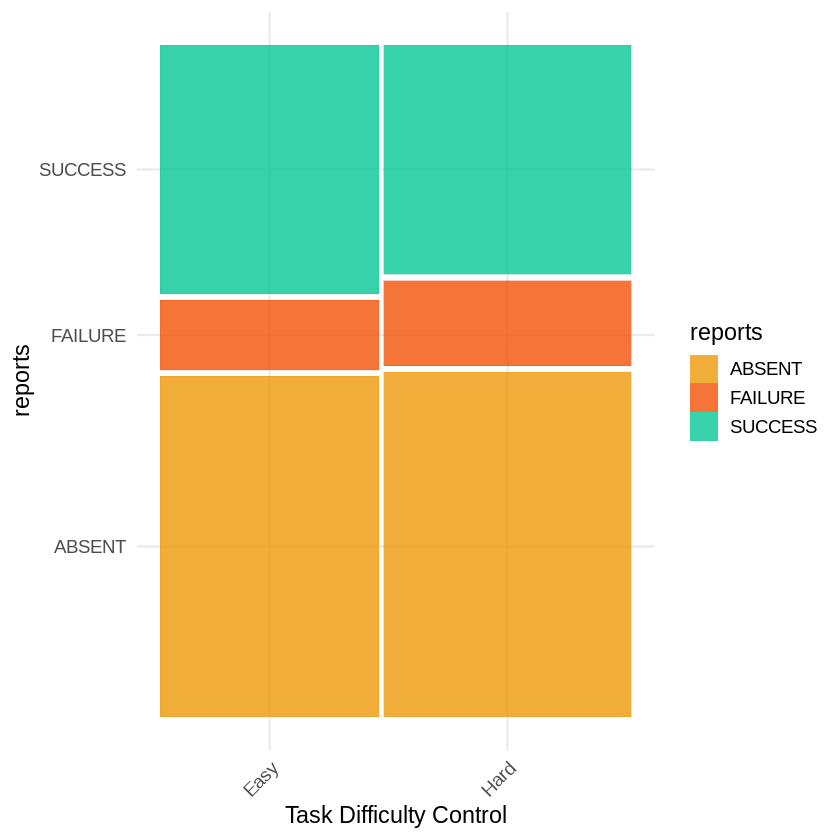}
        \caption{Proportion of success, failure, absent reports made under controlled difficulty condition.\label{fig:controlled_task_difficulty}}

    \end{subfigure}
    \hfill
    \begin{subtable}[t]{0.45\linewidth}
        \vskip 0pt
        \centering
        \begin{tabular}[t]{lll}
        \toprule
        \multicolumn{3}{c}{MODEL II}\\
        \midrule
        dependent $\rightarrow$ & \multirow{2}{*}{\emph{reported}} & \multirow{2}{*}{\emph{completed}} \\
        independent $\downarrow$ &  &  \\ \midrule
        \emph{day} & $-0.032^{***}$ & $-0.013.$  \\
        \emph{difficulty\_hard} & $-0.272$ & $-0.371$  \\ \midrule
         $R^2_m$ & $0.014$ & $0.007$ \\
         $R^2_c$ & $0.468$ & $0.478$\\ \bottomrule
        \end{tabular}
        \caption{Mixed-effects regression models for reports made and behavior successfully completed as a function of day and controlled difficulty measure in the app. p-value significance codes:  $^{***} 0.001, ^{**} 0.01, ^{*} 0.05, . 0.1$ }
         \label{tab:difficulty_condition}
    \end{subtable}
    \caption{Impact of controlled target behavior difficulty on behavioral compliance.}
\end{figure*}

Next, we study how difficulty of target behaviors impacts reporting and completion behaviors. Recall that we sorted participants randomly into \emph{easy} or \emph{hard} behavior difficulty group. During the behavior specification phase, the \emph{easy} group received a set of behaviors judged by the MTurk population to be low on the difficulty scale (described previously in Section \ref{sec:behavior-specification}). Similarly, the \emph{hard} group received a set of behaviors judged to be hard. Figure \ref{fig:controlled_task_difficulty} shows the proportion of success, failure, and absent reports made under both conditions. The participants made more successful reports in the \emph{easy} condition in comparison to the \emph{hard} condition. However, a mixed-effects regression for reported and completed dependent variables with day and behavior difficulty as independent variables show (Table \ref{tab:difficulty_condition}) that controlled behavior difficulty is not significant in predicting reporting or completing a behavior.

\begin{figure*}[h]
     \begin{subfigure}[b]{0.45\linewidth}

        \centering
        \includegraphics[width=1\textwidth]{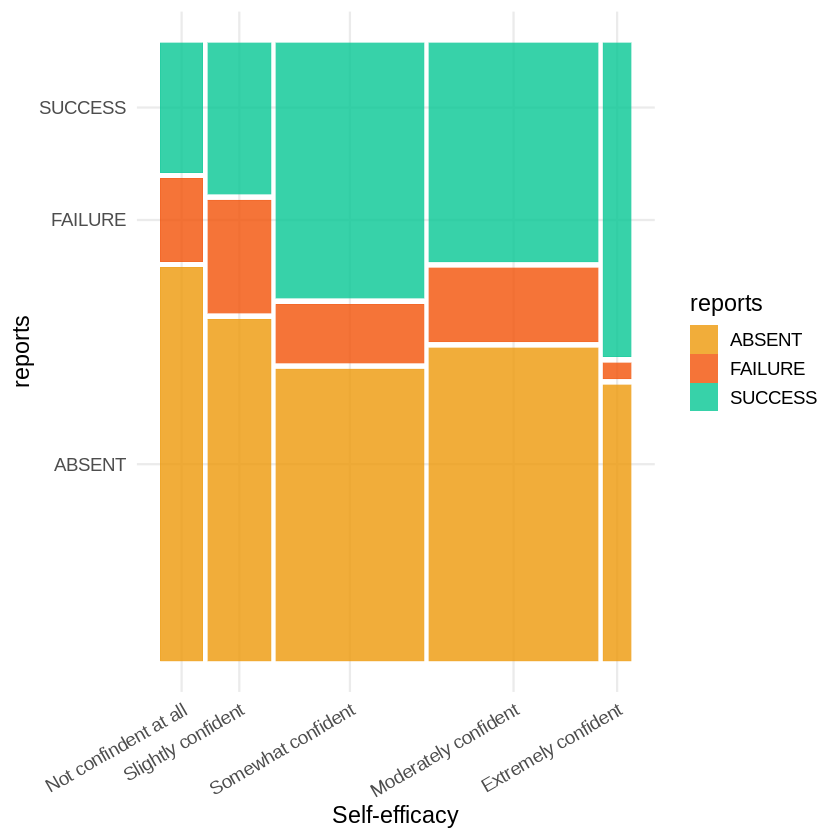}
        \caption{Proportion of success, failure, absent reports made under measured self-efficacy}
       \label{fig:app_self_efficacy}
    \end{subfigure}
    \hfill
    \begin{subtable}[b]{0.45\linewidth}

    \centering
    \begin{tabular}{lll}
    \toprule
        \multicolumn{3}{c}{MODEL III}\\
        \midrule
        dependent $\rightarrow$ & \multirow{2}{*}{\emph{reported}} & \multirow{2}{*}{\emph{completed}} \\
        independent $\downarrow$ &  &  \\ \midrule
        \emph{day} & $-0.032^{***}$ & $-0.013.$  \\
        \emph{self-efficacy} & $-0.216$ & $0.469^*$  \\ \midrule
        $R^2_m$ & $0.019$ & $0.039$ \\
        $R^2_c$ & $0.467$ & $0.482$\\\bottomrule
    \end{tabular}
    \vfill
    \caption{Mixed-effects regression models for reports made and behavior successfully completed as a function of self-reported difficulty. p-value significance codes:  $^{***} 0.001, ^{**} 0.01, ^{*} 0.05, . 0.1$ }
    \label{tab:difficulty_reported}
    \end{subtable}
    \caption{Impact of measured self-efficacy on behavioral compliance.}
\end{figure*}

During the behavior specification phase, each participant provided a self-efficacy evaluation of the target behavior they selected. Figure \ref{fig:app_self_efficacy} shows the distribution of success, failure, and absent reports made under varying levels of measured self-efficacy. We see that participants who selected behaviors that they were not confident about had more absent and failure reports than participants who selected behaviors they were confident about. Model III in Table \ref{tab:difficulty_reported} show the results from mixed-effects regressions for \emph{reported} and \emph{completed} dependent variables with day and measured self-efficacy as independent variables. We can see that that while self-efficacy doesn't affect the probability of reporting, it is significantly predictive of a person completing the target behavior they selected.

These findings support \ref{hypothesis:difficulty}, baseline ease of target behaviors positively influences behavior compliance. In particular, there is a strong individual component underlying the influence of behavior difficulty on compliance. How difficult a behavior is as evaluated by a large population in general does not influence how successful a person is performing a behavior as much as how difficult they thinks a behavior. Consequently, an mHealth system must be personalized
to each individual, recommending behaviors that the person believes they can do to ensure repetitive success.

\subsubsection{Reminders}
\label{sec:analysis_reminders}
Now we study the impact of reminding schedule on reporting and behavior compliance. Recall that participants who received reminders were randomly sorted into $4$ reminding schedules: $7$ or $14$ reminders each with \emph{uniform} or \emph{massed} distribution. Of $516$ reminders sent, $167$ were acknowledged while they were active. For the following analyses, we use acknowledgement as an indicator that the reminder sent actually was seen by the participant and may have influenced their behavior. If the reminder was not acknowledged, it is unlikely to have had any impact on the participant's behavior.

\begin{figure*}[h]
    \centering
    \begin{subfigure}[t]{0.44\linewidth}
    \vskip 0pt
        \centering
        \includegraphics[width=1\textwidth]{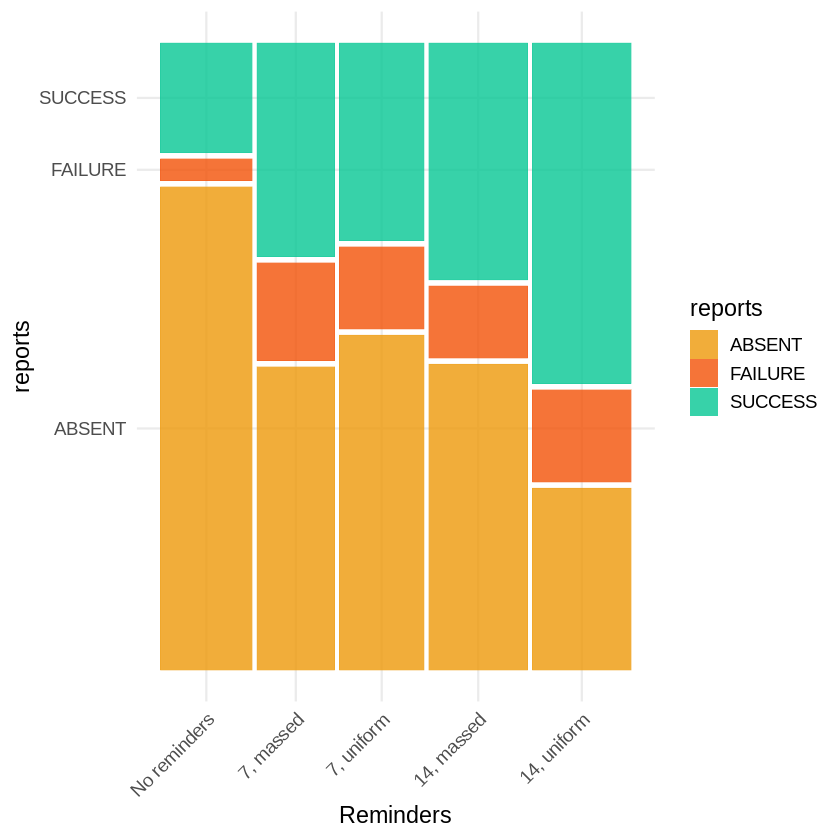}
        \caption{Proportion of success, failure, absent reports made under controlled reminder count and distribution}
        \label{fig:app_reminders}
    \end{subfigure}
    \begin{subtable}[t]{0.45\linewidth}
    \vskip 0pt
    \centering
    \begin{tabular}{lll}
    \toprule
    \multicolumn{3}{c}{MODEL IV}\\
    \toprule
    dependent $\rightarrow$ & \multirow{2}{*}{\emph{reported}} & \multirow{2}{*}{\emph{completed}} \\
    independent $\downarrow$ &  &  \\ \midrule
    \emph{day} & $-0.032^{***}$ & $-0.013.$  \\
    \emph{reminder-7} & $1.753^{**}$ & $1.529^{*}$  \\
    \emph{reminder-14} & $2.532^{***}$ & $2.398^{***}$  \\
    \midrule
    $R^2_m$ & $0.151$ & $0.124$ \\
    $R^2_c$ & $0.474$ & $0.491$\\ \bottomrule
    \end{tabular}
    \caption{Mixed-effects regression models for reports made and behavior successfully completed as a function of self-reported difficulty. p-value significance codes:  $^{***} 0.001, ^{**} 0.01, ^{*} 0.05, . 0.1$}
    \label{tab:reminders}
    \end{subtable}
    \caption{Mixed-effects regression models for reports made and behavior successfully completed as a function of self-reported difficulty. p-value significance codes:  $^{***} 0.001, ^{**} 0.01, ^{*} 0.05, . 0.1$}
  \label{fig:reminder}
\end{figure*}

Figure \ref{fig:app_reminders} shows the distribution of reports made under various reminder conditions. We see that more reports were made when participants were given reminders and a higher number of reminders resulted in a higher number of reports. Similar trend was observed in successful completion of selected target behavior. This finding is validated by the mixed-effects regression model in Table \ref{tab:reminders} in which we see that the impact of reminders is statistically significant. The impact of reminder distribution isn't immediately clear - in reminder count $n=7$ condition, massed distribution results in better compliance and higher number of reports than the uniform one, however the direction is reversed in $n=14$. Further analysis using mixed-effects modeling revealed that distribution wasn't a significant factor of reporting or compliance.

These results support \ref{hypothesis:frequency} that frequency of context reminders positively influences behavior compliance. Our CMC-based model of behavior suggests that associating behavior with contextual cues helps in improving compliance even when there is no explicit reminder - i.e, the presence of the associated context itself reminds the participant of a target behavior. To test this, we performed the following analyses on subset of data that includes only those days when the reminder was not acknowledged by the participant. Table \ref{tab:not_reminder_days} summarizes the impact of controlled reminder conditions on behavior on days when no reminders were acknowledged. We see that being reminded significantly improves reporting and compliance even on days when a participant doesn't read the reminder. This is evidence for our hypothesis that strengthening the association between context and behavior may improve overall compliance without there being a need for specific reminders - i.e, a habit is being built.

Finally, Table \ref{tab:reminder_count_impact} depicts the impact of reminders that were seen and acknowledged by participants. Here, the independent variable is not the experimental condition that the participant was assigned, but the number of reminders they have acknowledged during the experiment. As predicted by the CMC, a higher number of reminders leads to a stronger association between the context and the behavior which results in better reporting and compliance as evidenced in Table \ref{tab:reminders}.

\begin{figure*}[h]
    \centering
    \begin{subtable}[t]{0.45\linewidth}
    \vskip 0pt
    \centering
    \begin{tabular}{lll}
    \toprule
    \multicolumn{3}{c}{MODEL V}\\
    \toprule
    dependent $\rightarrow$ & \multirow{2}{*}{\emph{reported}} & \multirow{2}{*}{\emph{completed}} \\
    independent $\downarrow$ &  &  \\ \midrule
    \emph{day} & $-0.034^{***}$ & $-0.011$  \\
    \emph{reminder-7} & $1.779^{**}$ & $1.510^{*}$  \\
    \emph{reminder-14} & $2.560^{***}$ & $2.398^{***}$  \\
    \midrule
    $R^2_m$ & $0.162$ & $0.124$ \\
    $R^2_c$ & $0.474$ & $0.491$\\ \bottomrule
    \end{tabular}
    \caption{Mixed-effects regression models for reports made and behavior successfully completed as a function of controlled number of reminders on days when reminders were not acknowledged. p-value significance codes:  $*** 0.001, ** 0.01, * 0.05, . 0.1$}
    \label{tab:not_reminder_days}
    \end{subtable}
    \hfill
    \begin{subtable}[t]{0.45\linewidth}
    \vskip 0pt
    \centering
    \begin{tabular}{lll}
    \toprule
    \multicolumn{3}{c}{MODEL VI}\\
    \toprule
    dependent $\rightarrow$ & \multirow{2}{*}{\emph{reported}} & \multirow{2}{*}{\emph{completed}} \\
    independent $\downarrow$ &  &  \\ \midrule
    \emph{day} & $-0.057^{***}$ & $-0.036^{***}$  \\
    \emph{\# reminders} & $0.290^{***}$ & $0.239^{***}$  \\
    \midrule
    $R^2_m$ & $0.059$ & $0.033$ \\
    $R^2_c$ & $0.422$ & $0.439$\\ \bottomrule
    \end{tabular}
    \caption{Mixed-effects regression models for reports made and behavior successfully completed on days when the reminders were not acknowledged as a function of total number of reminders seen until then. p-value significance codes:  $*** 0.001, ** 0.01, * 0.05, . 0.1$}
    \label{tab:reminder_count_impact}
    \end{subtable}
    \caption{Mixed-effects regression models for reports made and behavior successfully completed as a function of self-reported difficulty. p-value significance codes:  $*** 0.001, ** 0.01, * 0.05, . 0.1$}
  \label{fig:reminder2}
\end{figure*}

\subsubsection{Difficulty and Reminders}
\label{sec:analysis_difficulty_reminders}
\begin{figure*}[h]
    \centering
    \begin{subfigure}[t]{0.45\linewidth}
    \vskip 0pt
        \centering
        \includegraphics[width=1\textwidth]{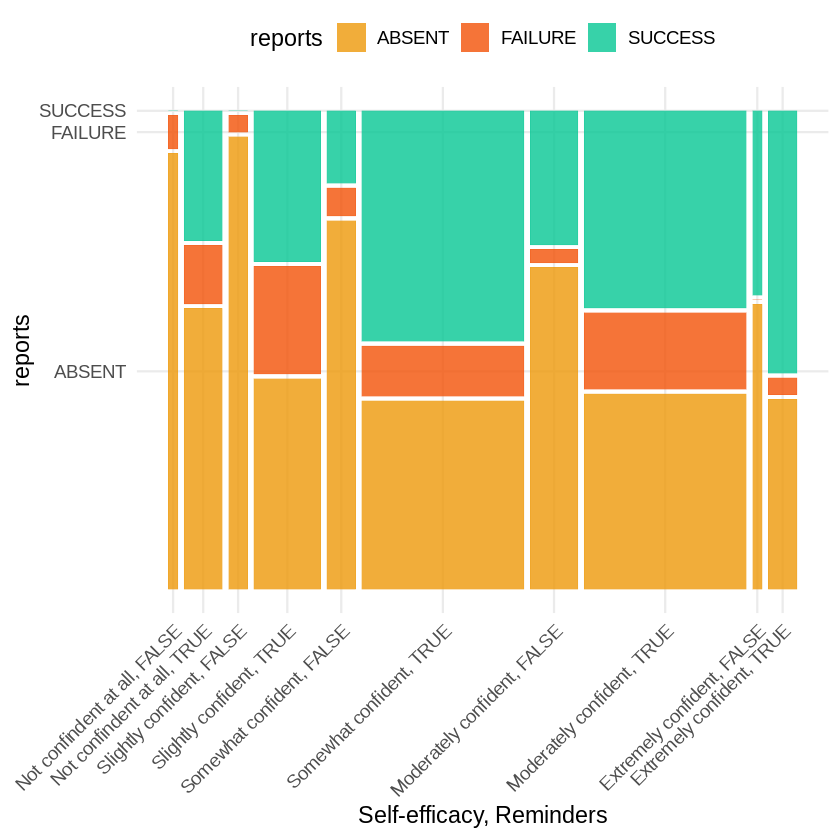}
        \caption{Proportion of success, failure, absent reports made under controlled reminder count and distribution}
        \label{fig:reported_self-efficacy_controlled_reminders}
    \end{subfigure}
    \hfill
    \begin{subtable}[t]{0.45\linewidth}
    \vskip 0pt
    \centering
    \begin{tabular}{lll}
    \toprule
    \multicolumn{3}{c}{MODEL VII}\\
    \toprule
    dependent $\rightarrow$ & \multirow{2}{*}{\emph{reported}} & \multirow{2}{*}{\emph{completed}} \\
    independent $\downarrow$ &  &  \\ \midrule
    \emph{day} & $-0.032^{***}$ & $-0.013.$  \\
    \emph{self-efficacy} & $0.252$ & $0.504^{*}$\\
    \emph{reminder-7} & $1.798^{***}$ & $1.653^{**}$  \\
    \emph{reminder-14} & $2.546^{***}$ & $2.471^{***}$  \\
    \midrule
    $R^2_m$ & $0.163$ & $0.168$ \\
    $R^2_c$ & $0.472$ & $0.497$\\ \bottomrule
    \end{tabular}
    \caption{Mixed-effects regression models for reports made and behavior successfully completed as a function of self-reported difficulty. p-value significance codes:  $^{***} 0.001, ^{**} 0.01, ^{*} 0.05, . 0.1$}
    \label{tab:reported_self-efficacy_controlled_reminders}
    \end{subtable}
    \caption{Mixed-effects regression models for reports made and behavior successfully completed as a function of self-reported difficulty. p-value significance codes:  $*** 0.001, ** 0.01, * 0.05, . 0.1$}
  \label{fig:reminder}
\end{figure*}

CMC predicts that difficulty and reminders will have an additive effect on behavior compliance they influence different pathways (\ref{hypothesis:reminding-additive}). Figure \ref{fig:reported_self-efficacy_controlled_reminders} provides some evidence for this hypothesis. We see that reminders positively impact success reports for all levels of self-efficacy. Mixed-effects regression models in Table \ref{tab:reported_self-efficacy_controlled_reminders} shows that the impact of reminder conditions is significant and influences completion above and beyond what is explained by self-efficacy measurement.

We can further see that the presence of reminders brings up the number of success reports to a greater extent in behaviors that are judged to be \emph{somewhat confident in}, \emph{moderately confident in}, and behaviors judged to be \emph{extremely confident in}. This result suggests an adaptive coaching strategy where if the coach wants to encourage a trainee to try relatively harder behaviors, it should send reminders more often to ensure the trainee makes an attempt to performing those. Reminders can serve as useful scaffolds until the trainee becomes used to performing those behaviors. Another interesting observation is that the number of failure reports are also higher in the presence of reminders as against absent ones, regardless of level of behavioral difficulty. This finding is important because adaptive coaching algorithms can greatly benefit from a trainee's reported failure rather than having to guess or estimate behavioral compliance when a report is absent. Therefore, this impact of reminders on increasing participants' reports can be useful in designing adaptive computer coaches for health behavior change.

\subsubsection{Judgments and Attitudes}
\label{sec:analysis_judgments}
\begin{table}[t]
    \vskip 0pt
    \centering
    \begin{tabular}{lcccc}
    \toprule
    \multicolumn{5}{c}{MODEL VII}\\
    \toprule
    dependent $\rightarrow$ & \multirow{2}{*}{\emph{difficulty}} & \multirow{2}{*}{\emph{self-efficacy}} & \multirow{2}{*}{\emph{affective attitude}}& \multirow{2}{*}{\emph{instrumental attitude}} \\
    independent $\downarrow$ &  &  \\ \midrule
    \emph{\# successes} & $-0.039^{***}$ & $0.043^{***}$ & $0.046^{***}$ & $0.037^{***}$\\
    \emph{\# absents} & $0.026^*$ & $-0.024^{*}$ & $0.004$ & $-0.005$\\
    \emph{\# failures} & $0.0680^{**}$ & $-0.067^{***}$ & $-0.077^{***}$ & $-0.064^{***}$\\
    \midrule
    $R^2_m$ & $0.052$ & $0.072$ & $0.064$ & $0.058$\\
    $R^2_c$ & $0.229$ & $0.550$ & $0.561$ & $0.675$\\ \bottomrule
    \end{tabular}
    \caption{Mixed-effects regression models for reports made and behavior successfully completed as a function of self-reported difficulty. p-value significance codes:  $^{***} 0.001, ^{**} 0.01, ^{*} 0.05, . 0.1$}
    \label{tab:valued_judgement1}
\end{table}

Recall that CMC posits that previous experience with a target behavior influences how it is evaluated amongst behaviors competing for resources. If a trainee highly successful at a target behavior, they will value the behavior highly. And, if the trainees value certain behaviors highly, they will execute them when opportune situations arise with high likelihood. Therefore, for sustainable behavioral compliance, it is important to understand how a trainee's judgment of behaviors evolve as they gain experience performing them. We evaluate this hypothesis - \ref{hypothesis:value} in this section.

First, we study how successful experience with a target behavior influences its evaluation. For this analysis, our dependent variables are daily measurements of \emph{difficulty}, \emph{self-efficacy}, \emph{affective attitude}, and \emph{instrumental attitude}. For each dependent variable, we developed a mixed-effects linear regression model. Independent variables in this analysis include number of success, failure, and absent reports made until the judgement was measured. For example, for judgement measurement on day $10$, we counted how many success, failure, and absent reports were made on days $1-9$. Table \ref{tab:valued_judgement1} summarizes the results of these models. We can see that the \emph{difficulty} value judgment is positively correlated with number of failures experienced and inversely with the number of successes experienced. We observe an opposite pattern with \emph{self-efficacy}, \emph{affective attitude}, and \emph{instrumental attitude} judgments. This finding demonstrates that value judgments are influenced by a trainee's experiences of practicing a target behavior. Consequently, to build a sustainable habit, the coach must help the trainee have successful experiences with the target behavior by gradually adapting the difficulty of the behavior.

Next, we analyze if these judgments impact success at executing the target behavior. We developed mixed-effects logistic models with success, failure, and absent report as dependent variables. Indepedent variables were computed as a moving average of measured value judgements with a window of $3$ days. We used a moving window of $3$ because prior work \cite{pirolli2016computational} suggests that there is a significant recency effect because older judgments may not be easily accessible in episodic memory and therefore, have limited impact on behavior. Figure \ref{tab:value_judgement2} shows the coefficients from the logistic model estimated from our data. We see that difficulty and self-efficacy are predictive of successes and failures, but instrumental and affective attitude aren't. Absent reports were not predicted by any judgments. From these analyses, we conclude that \ref{hypothesis:value} is supported in our dataset.

\begin{table}[t]
    \vskip 0pt
    \centering
    \begin{tabular}{lccc}
    \toprule
    \multicolumn{4}{c}{MODEL VIII}\\
    \toprule
    dependent $\rightarrow$ & \multirow{2}{*}{\emph{success}} & \multirow{2}{*}{\emph{failure}} & \multirow{2}{*}{\emph{absent}} \\
    independent $\downarrow$ &  &  \\ \midrule
    \emph{average difficulty} & $-0.475^{***}$ & $0.678^{***}$ & $0.040$\\
    \emph{average self-efficacy} & $0.180.$ & $-0.336^{**}$ & $0.024$\\
    \emph{average affective attitude} & $0.039$ & $-0.184.$ & $0.051$\\
    \emph{average instrumental attitude} & $-0.165$ & $0.110$ & $0.039$\\
    \midrule
    $R^2_m$ & $0.080$ & $0.206$ & $0.002$ \\
    $R^2_c$ & $0.253$ & $0.303$ & $0.177$ \\ \bottomrule
    \end{tabular}
    \caption{Mixed-effects regression models for reports made and behavior successfully completed as a function of self-reported difficulty. p-value significance codes:  $^{***} 0.001, ^{**} 0.01, ^{*} 0.05, . 0.1$}
    \label{tab:value_judgement2}
\end{table}

\subsubsection{Pre and Post Survey Measures}
\label{sec:survey}
 In this study, we did not find any statistically significant changes in any of these outcomes measured post- compared to pre-intervention. This may be in part due to the short duration of the intervention i.e., $4$ weeks, which may have been an insufficient dose to trigger noticeable changes. We argue that this supports the need to further design and study behavior change systems such as PARCcoach, which can measure and track day-to-day dynamics of health behaviors at a finer granularity.

\subsubsection{Usability of PARCcoach}
\label{sec:usability}
\begin{figure}[h]
    \centering
    \includegraphics[width=0.5\textwidth]{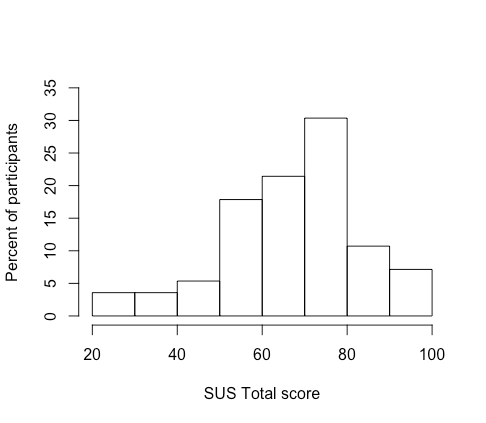}
    \caption{Histogram of SUS composite score}
    \label{fig:sus}
\end{figure}

In order to characterize and benchmark the usability of PARCcoach, we administered the $10$-item System Usability Scale (SUS) \cite{brooke1996sus} as part of the post-survey that produced a total composite score on a $100$-point scale interpreted as percentile ranking i.e., \emph{perceived usability}. Overall, participants demonstrated acceptable usability with a median composite SUS score of $70$ across the $60$ participants who completed the study (composite SUS mean $68.35 \pm 15.34$). Figure \ref{fig:sus} shows the distribution of SUS composite scores in our study as a histogram indicating that more than $50$\% of participants had calculated scores higher than $60$. Based on \cite{sauro2011practical}, this would put our system at a ranking of \emph{B-}, which is reasonable for a first of its kind.

We also asked participants to report qualitative feedback on the system in the post-survey by answering open-ended questions on what they ``liked" and ``did not like" about the app. Participants most frequently reported that they ``liked'' that the app was easy to use and kept them accountable to follow their health goals. However, they also indicated that they ``did not like'' the fact that the app was too simple, did not ``learn'' from their reports and adapt to that, and the fact that they could not go back to report on a previous day's goal. It is interesting to note that participants start to expect personalized, adaptive features based on the current digital ecosystem regardless of the context. On the other hand, our design choice to prevent ``back-reporting'' i.e., changing any previous day's goal report was clearly bothersome to participants. However, back-reporting can compromise accuracy of goal reporting as shown in prior work \cite{du2016group}. Our observations suggest that allowing ``back-reports'' is a design choice mHealth developers must carefully make, balancing need for accuracy with users' expectations.

\subsubsection{Summary of Findings}
From our $4$-week deployment study of PARCcoach, we can conclude the following:
\begin{enumerate}
    \item Participants were able to use PARCcoach to select a target behavior and maintain a record of their successes and failures at achieving it every day.
    \item Participants who selected target behaviors that they were confident in (behaviors were not difficult or hard) attempting were more likely to comply with it and consequently more likely to build a healthy habit. We observed a strong personal component and consequently, an ideal adaptive coach should be cognizant of difficulty as it pertains to each individual trainee as per desideratum \ref{desiderata:2}.
    \item CMC-based view suggests a novel strategy of reminding that relies on the CMC to retrieve the right goal at the right time. This is ensured by creating associations between elements of the environment context with a target behavior and strengthening the associations through reminders. We observed that such reminders are useful in improving behavioral compliance and support desiderata \ref{desiderata:3} and \ref{desiderata:4}.
    \item CMC-based view posits that difficulty of target behavior and reminding have additive effects on behavior compliance. This is because difficulty and reminding affect different aspects of the CMC. This was supported by our data.
    \item Participants' value judgements were influenced by their successes and failures at attempting the target behavior. Further, value judgements were predictive of behavior compliance. This finding supports desideratum \ref{desiderata:5}.
\end{enumerate}
Finally, while this paper does not extensively evaluate the common model of cognition itself, it provides evidence that the CMC is a useful framework to explain behavior change and to motivate design desiderata for an intelligent health coach.

\section{Conclusions and Future Work}
Health behaviors account for an estimated of $60\%$ of the risks associated with chronic illnesses such as diabetes and cardiovascular disease. As the at-risk population grows around the world, the challenge of developing and disseminating effective methods for improving health behaviors is becoming critically important. A wide range of strategies (or \emph{interventions}) have been investigated for promoting better health behaviors \cite{KahnTheServices} including informational approaches to change knowledge and attitudes toward health, behavioral counseling approaches to teach and maintain health-related skills, creating social environments to support health behavior change, and environmental and policy changes to provide resources that enable good health behaviors. Of these, individual counseling received in personal meetings \cite{KarenB.Eden2002DoesEvidence} or over telephone \cite{Eakin2007TelephoneReview} have been shown to be very effective in promoting behavior change. Although useful, such counseling is very resource-intensive, in terms of both training necessary personnel and delivering to a large population.

To make such counseling pervasive and cost-effective, we are motivated to develop an interactive, intelligent agent that can reside on a mobile device and provide counseling in a manner similar to a human coach. Human health coaches rely on a collection of behavior-change strategies to motivate their trainees to make progress toward their health goals. To design mHealth systems that can opportunistically employ various behavior change strategies, it is important to develop a causal model of human behavior as well as characterize the space of adaptation in coach-trainee interaction. In this paper, we leverage the common model of cognition \cite{laird2017standard} as a framework to situate and unify several behavior change theories - goal setting, implementation intentions, reminding, and judgements and attitudes. We demonstrate that this view is useful in designing a comprehensive, interactive mHealth system that employs these theories to guide computer-human interaction. Our experimental evaluation shows promising results and shows that integration of these strategies indeed is useful in promoting behavior change. We demonstrate how information collected during implementation intention setting can be used to greatly personalize remindings. We show that incorporating individual measures of behavior difficulty and reminders increases behavior compliance above and beyond what can be achieved through goal setting alone.

While our initial findings are exciting, we also uncovered several design challenges in developing, administering, and evaluating such interventions over prolonged periods of time. To begin with, future investigations are needed to inform design of these systems that are more user-friendly in order to match user expectations for \emph{adaptive} recommendations based on previous goal completion. Future work is needed to incorporate artificial intelligence-based recommendations that are built on individualized models of behavior. Next, there needs to be additional work in characterizing day-to-day behavioral dynamics to identify effectiveness of interventions to overcome the lower sensitivity of validated outcome measures and survey instruments in identifying subtle changes in health behaviors over shorter intervention windows. Moreover, system and intervention design needs to be able to better account for erroneous self-reported, subjective outcome measures. Future work in these particular areas can help significantly advance the design and larger scale deployment of mHealth-based behavior change programs to promote better health and reduce chronic disease burden overall.

\section{Acknowledgments}
The author would like to thank Peter Pirolli and Anusha Venkatakrishnan for their contributions to the design of PARCCoach and the study described in this paper, Les Nelson and Michael Silva for engineering the system, and Anusha Venkatakrishnan and Aaron Springer for recruiting participants and running the study. The author additionally appreciates Jacqui LeBlanc for helping us determine the relevant habits and behaviors to study. The author is thankful to Shekhar Mittal for supporting the data analysis presented in this paper. The development of PARCCoach and the study presented here was supported in part by the National Science Foundation under Grant No. $1346066$ to Peter Pirolli and by Xerox Corporation.
\bibliographystyle{ACM-Reference-Format}
\bibliography{references}
\end{document}